 \documentclass[onecolumn]{aastex631}

\usepackage{graphicx}	
\usepackage{amsmath}	
\usepackage{xcolor}
\usepackage{soul}
\usepackage{float}
\usepackage{placeins}  
\usepackage{array,makecell}
\usepackage{dcolumn}
\newcolumntype{d}[1]{D{.}{.}{#1}}
\usepackage{longtable}
\usepackage{array}

\newcommand{\mcellbig}[2][1.10]{\scalebox{#1}{$\displaystyle #2$}}

\graphicspath{{./}{figures/}}

\shorttitle{Shape-Preserving Evolution of the UV QLF}
\shortauthors{Wang et al.}

\begin{document}

\title{Shape-Preserving Evolution of the Global Ultraviolet Quasar Luminosity Function to $z\simeq7.5$}

\author[0009-0005-1617-2442]{Wenjie Wang}
\affiliation{Department of Physics, School of Physics and Electronics, Hunan Normal University, Changsha 410081, People's Republic of China}
\affiliation{Key Laboratory of Low-dimensional Quantum Structures and Quantum Control, Hunan Normal University, Changsha 410081, People's Republic of China}
\affiliation{Hunan Research Center of the Basic Discipline for Quantum Effects and Quantum Technologies, Hunan Normal University, Changsha 410081, People's Republic of China}

\author[0000-0001-6861-0022]{Zunli Yuan}
\affiliation{Department of Physics, School of Physics and Electronics, Hunan Normal University, Changsha 410081, People's Republic of China}
\affiliation{Key Laboratory of Low-dimensional Quantum Structures and Quantum Control, Hunan Normal University, Changsha 410081, People's Republic of China}
\affiliation{Hunan Research Center of the Basic Discipline for Quantum Effects and Quantum Technologies, Hunan Normal University, Changsha 410081, People's Republic of China}

\author{Yu Luo}
\affiliation{Department of Physics, School of Physics and Electronics, Hunan Normal University, Changsha 410081, People's Republic of China}
\affiliation{Key Laboratory of Low-dimensional Quantum Structures and Quantum Control, Hunan Normal University, Changsha 410081, People's Republic of China}
\affiliation{Hunan Research Center of the Basic Discipline for Quantum Effects and Quantum Technologies, Hunan Normal University, Changsha 410081, People's Republic of China}

\author{Longhua Qin}
\affiliation{Department of Physics, Yuxi Normal University, Yuxi, Yunnan, 653100, China}

\correspondingauthor{Zunli Yuan}
\email{yzl@hunnu.edu.cn}

\begin{abstract}

We present a global unbinned-likelihood analysis of the rest-frame ultraviolet (UV) quasar luminosity function (QLF) over $0.1\le z\le7.5$, using a final analysis sample of 70,960 Type~1 quasars selected from the homogenized compilation of Kulkarni et al. (2019).
We test a shape-preserving luminosity and density evolution (LADE) framework, in which the local luminosity function (LF) shape is fixed and the redshift evolution is described by separate density- and luminosity-evolution factors.
For a double-power-law (DPL) local LF, we search 81 LADE models built from nine density-evolution and nine luminosity-evolution functions, and compare them with two flexible double-power-law (FDPL) reference models.
The fiducial shape-preserving LADE model gives the lowest Akaike information criterion (AIC) and Bayesian information criterion (BIC) values among the main models considered, and 12 DPL-based LADE models have lower AIC and BIC values than both FDPL reference models.
Repeating the model-grid analysis with a modified-Schechter local LF gives the same preferred evolutionary structure, indicating that the result is not driven only by the assumed local LF shape.
The fitted evolution functions further show that the luminosity-evolution component is the more stable part of the LADE decomposition: it rises rapidly to $z\simeq2$--3 and then flattens or slowly declines.
The density-evolution component is more model dependent, but the preferred LADE models consistently require a declining effective normalization toward high redshift.
Taken together, we conclude that a shape-preserving framework offers a statistically efficient and compact empirical description of the global UV QLF.

\end{abstract}


\section{Introduction}
\label{sec:introduction}

The luminosity function (LF) of active galactic nuclei (AGN) and its evolution provide key empirical constraints on the build-up of supermassive black holes and models of galaxy formation
\citep{1982MNRAS.200..115S,2000ApJ...531...42H,2004ApJ...602..603Y,2007ApJ...654..731H,2008ApJ...679..118S,
2009A&A...493...55E,2010MNRAS.401.2531A,2013ApJ...773...14R,2014ApJ...787...73D,2006ApJ...650...42L,
2008MNRAS.385.1846M}.
In the rest-frame ultraviolet (UV), the quasar luminosity function (QLF), $\phi(z,M_{1450})$, provides a direct census of unobscured accretion.
With appropriate bolometric and obscuration corrections, integrating the LF over luminosity and redshift can be related to the growth of the supermassive black hole (SMBH) mass density \citep[e.g.,][]{1982MNRAS.200..115S,2002MNRAS.335..965Y}.
The UV QLF is also an important input to models of the metagalactic ionizing background and quasar contributions to hydrogen reionization and the post-reionization UV background \citep{2012ApJ...746..125H,Giallongo2015,2017ApJ...847L..15O,2019MNRAS.485...47P,2019MNRAS.488.1035K}.
A robust global description of $\phi(z,M_{1450})$ over cosmic time is therefore essential.

Over the past two decades, wide-field imaging and systematic spectroscopic follow-up have enabled increasingly precise QLF measurements across a broad dynamic range in luminosity and redshift.
At low to intermediate redshift, the constraints are dominated by large-area optical samples with high spectroscopic completeness and well-characterized selection functions \citep{2000MNRAS.317.1014B,2004MNRAS.349.1397C,2009MNRAS.392...19C,2006AJ....131.2766R,2010AJ....139.2360S,2013ApJ...773...14R,2017A&A...597A..79P,2018A&A...613A..51P,2020ApJS..250....8L}.
At higher redshift, bright-end constraints are provided by wide-area color selections and optical+IR searches \citep{2001AJ....121...54F,2006AJ....132..117F,2013ApJ...768..105M,2016ApJ...833..222J,2016ApJ...829...33Y,2021ApJ...912..111B}, while the faint end relies increasingly on deep multiwavelength fields with intensive spectroscopy \citep{2011ApJ...728L..26G,Giallongo2015,2018PASJ...70S..34A,2018ApJ...869..150M,2017ApJ...847L..15O}.
At $z\gtrsim6$, the LF is constrained mainly by rare luminous quasars from wide-area optical and near-IR surveys \citep{2011Natur.474..616M,2015ApJ...801L..11V,2018Natur.553..473B,2019ApJ...884...30W,2019AJ....157..236Y}.

A persistent challenge for global QLF inference is the heterogeneity of these constraints.
Different surveys populate different regions of the $M_{1450}$--$z$ plane, and the inferred space densities depend sensitively on selection boundaries, completeness, and survey-specific systematics.
A homogeneous treatment of the input sample and selection functions is therefore crucial for likelihood-based global modeling.

A major step in this direction was provided by the homogenized UV Type~1 quasar compilation of \citet[][hereafter K19]{2019MNRAS.488.1035K}.
K19 placed multiple survey components on a uniform $M_{1450}$ system and provided survey-specific selection functions $f(z,M_{1450})$.
They also introduced a flexible global QLF parameterization in which the four double-power-law (DPL) parameters are allowed to evolve with redshift through Chebyshev polynomials.
This flexible double-power-law (FDPL) model provides an important empirical reference description of the global UV QLF, but it also raises a natural question: are all of these redshift-dependent degrees of freedom required by the data?

This motivates the model-comparison problem addressed in this work.
We ask whether the global UV QLF can instead be described by a shape-preserving luminosity and density evolution (LADE) framework, in which the local LF shape is fixed and the redshift evolution is represented by a density-evolution factor, $e_1(z)$, and a luminosity-evolution factor, $e_2(z)$.
Such a model is more restrictive than the FDPL parameterization because it does not allow the faint- and bright-end slopes to vary independently with redshift.
At the same time, LADE-type descriptions have been widely used in LF studies at other wavelengths, where luminosity and density evolution are factorized while the LF shape parameters are held fixed \citep[e.g.,][]{2010MNRAS.401.2531A,yuan2016mixture,yuan2017mixture,Novak_2017,2022ApJ...941...10V,2024A&A...683A.174W,2026arXiv260315449W,2026ApJ...997..176W}.
This makes the LADE framework a natural and physically constrained alternative for testing whether the observed UV QLF evolution can be described primarily by changes in normalization and characteristic luminosity, rather than by explicit redshift evolution of the LF shape.

In this work, we perform a global unbinned-likelihood analysis of the K19 compilation within this shape-preserving LADE framework.
Relative to the K19 fiducial global-fit configuration, we conservatively re-include the low-redshift interval $0.1\le z<0.6$ and a bright-end subset of the BOSS DR9 sample, using luminosity cuts designed to reduce sensitivity to the systematic regimes identified by K19.
We construct a grid of LADE models from candidate density- and luminosity-evolution functions, test both the conventional DPL local LF and the Saunders modified-Schechter local LF introduced by \citet{saunders199060}, and compare the preferred shape-preserving models with two K19-style FDPL reference models using the same final sample, likelihood formalism, and survey selection functions.
This setup provides a controlled statistical test of whether a compact, physically more restrictive shape-preserving description can reproduce the global rest-frame UV QLF over $0.1\le z\le7.5$.

This paper is organized as follows.
Section~\ref{sec:data} describes the parent data compilation, survey selection functions, and our conservative re-inclusion of the low-redshift and BOSS DR9 samples.
Section~\ref{sec:methods} presents the QLF definition and likelihood formalism, the shape-preserving LADE model construction, the two local LF choices, the FDPL reference models, and the model-selection criteria.
Section~\ref{sec:results} presents the best-fitting QLFs, the density--luminosity decomposition, and the cumulative number-density evolution implied by the preferred models.
Section~\ref{sec:discussion} discusses the empirical implications, robustness, model dependence, and limitations of the shape-preserving interpretation.
We summarize our main conclusions in Section~\ref{sec:conclusions}.

Throughout this work we adopt a flat $\Lambda$CDM cosmology with
$H_0=70~\mathrm{km\,s^{-1}\,Mpc^{-1}}$, $\Omega_{\rm m}=0.3$, and $\Omega_{\Lambda}=0.7$.
Distances and volumes are expressed in comoving units, and we denote comoving megaparsecs as cMpc.
All magnitudes are reported in the AB system \citep{1983ApJ...266..713O}.
We characterize the rest-frame UV output using the absolute magnitude at 1450~\AA, $M_{1450}$.

\begin{figure*}[htbp]
	\centering
	\includegraphics[width=0.9\textwidth]{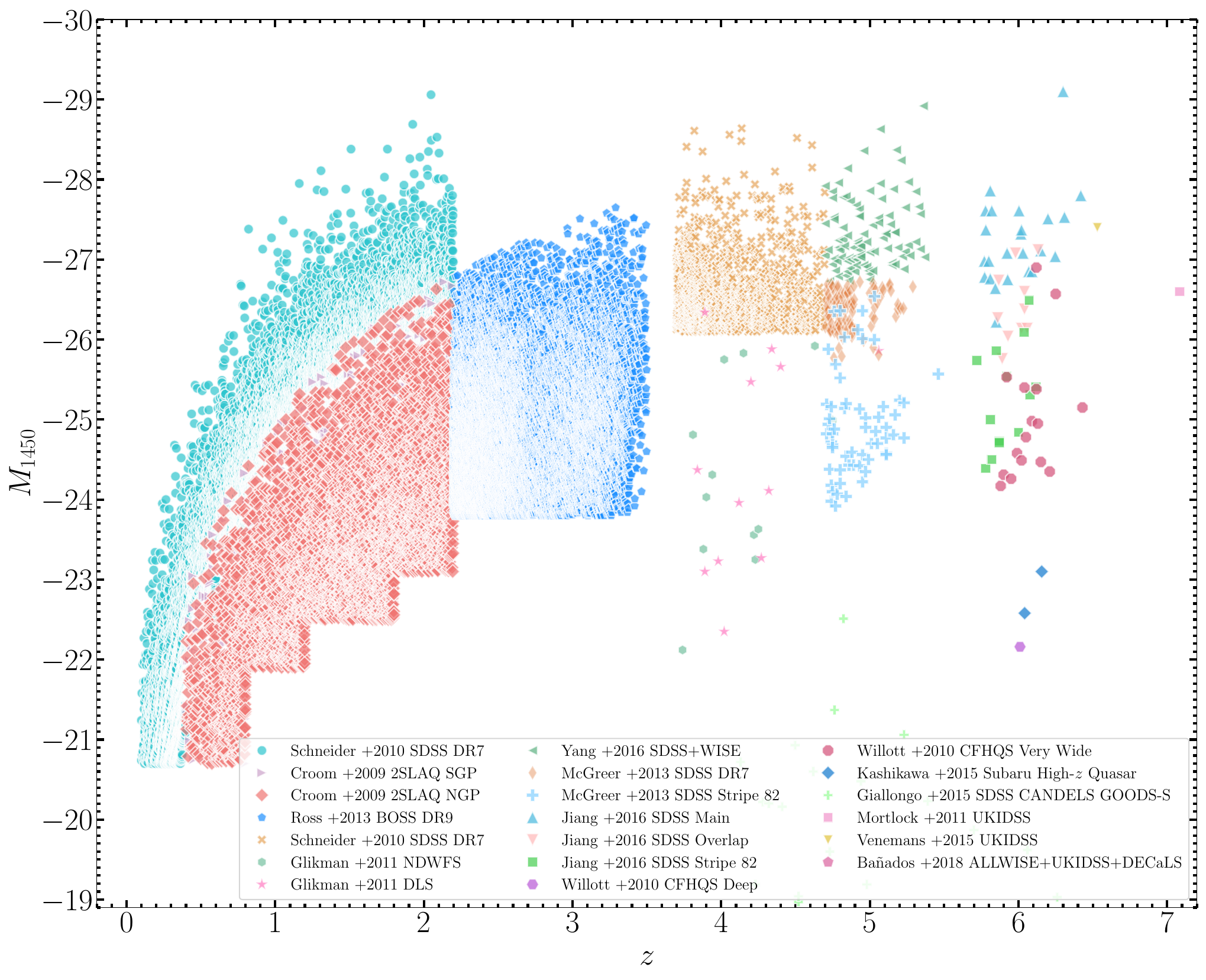}
	\caption{
    Distribution of the final analysis quasar sample in the plane of rest-frame UV absolute magnitude, $M_{1450}$, and redshift.
     Individual data points are color-coded according to their parent survey components from the K19 compilation, as indicated in the legend. Because the sample combines survey components with different depths, redshift windows, and selection functions, the point distribution should not be interpreted as a single flux-limited sample with one sharp faint-end boundary.
    }
	\label{fig:sample_data}
\end{figure*}

\begin{table*}[htbp]
\centering
\footnotesize
\caption{Quasar samples analyzed in this work.
The ``Final Number'' column gives the number of objects retained after applying the survey validation ranges, the conservative low-redshift cuts, and the BOSS bright-end cut used in the unbinned likelihood analysis.}
\hspace*{-1.1cm}
\begin{tabular}{lllrr}
\hline\hline
Survey & $z$ range & Reference & Original Number & Final Number \\
\hline
SDSS DR7                                   & $0.0$--$2.2$ & \citet{2010AJ....139.2360S} & 48,664 & 44,434 \\
2SLAQ SGP                                  & $0.4$--$2.2$ & \citet{2009MNRAS.392...19C} & 2,338  & 1,867  \\
2SLAQ NGP                                  & $0.4$--$2.2$ & \citet{2009MNRAS.392...19C} & 7,027  & 5,609  \\
BOSS DR9                                   & $2.2$--$3.5$ & \citet{2013ApJ...773...14R} & 23,301 & 17,629 \\
SDSS DR7                                   & $3.7$--$4.7$ & \citet{2010AJ....139.2360S} & 1,785  & 1,048  \\
NDWFS                                      & $3.6$--$5.2$ & \citet{2011ApJ...728L..26G} & 12    & 12    \\
DLS                                        & $3.8$--$5.3$ & \citet{2011ApJ...728L..26G} & 12    & 12    \\
SDSS+WISE                                  & $4.7$--$5.4$ & \citet{2016ApJ...829...33Y} & 99    & 99    \\
SDSS DR7                                   & $4.7$--$5.5$ & \citet{2013ApJ...768..105M} & 103   & 103   \\
Stripe 82                                  & $4.7$--$5.5$ & \citet{2013ApJ...768..105M} & 59    & 59    \\
SDSS main                                  & $5.7$--$6.5$ & \citet{2016ApJ...833..222J} & 24    & 24    \\
SDSS overlap                               & $5.7$--$6.5$ & \citet{2016ApJ...833..222J} & 10    & 10    \\
Stripe 82                                  & $5.7$--$6.5$ & \citet{2016ApJ...833..222J} & 13    & 13    \\
CFHQS Deep                                 & $5.8$--$6.6$ & \citet{2010AJ....139..906W} & 1     & 1     \\
CFHQS Very Wide                            & $5.8$--$6.6$ & \citet{2010AJ....139..906W} & 16    & 16    \\
Subaru High-$z$ Quasar                     & $5.8$--$6.5$ & \citet{2015ApJ...798...28K} & 2     & 2     \\
CANDELS GOODS-S                            & $4.0$--$6.5$ & \citet{Giallongo2015} & 19    & 19    \\
UKIDSS                                     & $6.5$--$7.4$ & \citet{2011Natur.474..616M} & 1 & 1 \\
UKIDSS                                     & $6.5$--$7.4$ & \citet{2015ApJ...801L..11V} & 1 & 1 \\
ALLWISE+UKIDSS+DECaLS                      & $6.5$--$7.4$ & \citet{2018Natur.553..473B} & 1     & 1     \\
Total &  &  &  83,488 & 70,960  \\
\hline
\hline
\end{tabular}
\label{tab:data}
\end{table*}

\section{Data and Sample Selection}
\label{sec:data}

\subsection{Parent sample}

This work is based on the homogenized rest-frame UV Type~1 quasar compilation of K19, which combines 20 survey components to provide broad area--depth coverage over $0\le z\le 7.5$ on a uniform $M_{1450}$ system.
The parent compilation contains 83,488 sources.
In Table~\ref{tab:data} we list the survey components used here and the object counts before and after applying our data-usage cuts; the final analysis sample contains 70,960 Type~1 quasars.
The resulting $M_{1450}$--$z$ distribution is shown in Figure~\ref{fig:sample_data}.

Following K19, the transformation from an observed selection-band magnitude $m$ to $M_{1450}$ is
\begin{equation}
M_{1450}(z) = m - 5\log_{10}\!\left[\frac{d_L(z)}{\mathrm{Mpc}}\right] - 25 - K_{m,1450}(z),
\end{equation}
where $d_L(z)$ is the luminosity distance and $K_{m,1450}(z)$ is the bandpass correction to 1450,\AA.
The K19 $K$-corrections use a combination of the \citet{2015MNRAS.449.4204L} stacked quasar spectrum at $\lambda<2500$~\AA~and the \citet{2001AJ....122..549V} quasar composite spectrum at longer wavelengths.
We adopt the homogenized $M_{1450}$ values and $K$-corrections from K19 throughout.

\subsection{Survey samples and selection functions}

At $z\lesssim2.2$, the K19 compilation is dominated by large optical spectroscopic surveys, primarily SDSS DR7 and 2SLAQ.
It then incorporates BOSS DR9 over $2.2\le z<3.5$, and at higher redshift is supplemented by a combination of wide-area bright-end samples and deep fields that constrain the faint end (see Table~\ref{tab:data}; also K19 Table~1 for component definitions, areas, and parent samples).
A defining feature of the compilation is the provision of survey-specific selection functions, $f(z,M_{1450})$, encoding photometric selection, imaging incompleteness, and, where applicable, spectroscopic targeting and redshift success.
We do not re-derive completeness corrections; instead, we adopt the homogenized $f(z,M_{1450})$ from K19 and incorporate them in the effective-volume normalization of our unbinned likelihood (Section~\ref{sec:methods}).

The sample distribution in Figure~\ref{fig:sample_data} should therefore not be interpreted as the selection boundary of a single flux-limited survey.
This is especially relevant at $z\gtrsim4$, where the compilation combines wide-area bright-end searches and deep-field samples with different sky coverages, redshift windows, and selection functions.
The resulting point distribution is expected to appear sparse and heterogeneous in the $M_{1450}$--$z$ plane.
In the likelihood analysis, this heterogeneity is accounted for through the survey-specific selection functions and effective-volume integrals, rather than through a single apparent magnitude limit.

We next describe how these survey components are used in the present analysis.
For the global analysis, we use the K19 homogenized catalogue and selection functions as our parent data set, but adopt a different data-usage strategy in the two redshift regimes where K19 identified the strongest systematic concerns.
K19 excluded the BOSS DR9 sample from their global evolutionary fits because the binned DPL fits over $2.2\le z<3.5$ showed short-redshift-scale scatter, a discontinuity relative to SDSS+2SLAQ at $z=2.2$, and an apparently rapid evolution of the faint-end slope that they argued was unlikely to be physical.
They attributed this behavior to the systematics limit of the large BOSS sample, whose fixed selection function depends on assumptions about the quasar spectral energy distribution (SED), intergalactic medium (IGM) absorption, and photometric errors near the targeted magnitude limit.
K19 also excluded $z<0.6$ quasars because the low-redshift faint-end slope shows a sharp increase, attributed to residual uncertainties in host-galaxy light corrections and potentially missed AGN in extended sources.

These concerns are directly relevant to our analysis, but our use of the data is different from that of K19.
K19 used these diagnostics to decide which redshift bins should be excluded from a flexible redshift-dependent DPL evolution fit.
In contrast, our aim is to test whether the global QLF can be described by a shape-preserving LADE model.
In such a framework, the low-redshift data provide an important local anchor for the fixed LF shape and normalization, while the BOSS redshift interval provides substantial leverage near the peak epoch of quasar activity.
We therefore re-include these regimes only after imposing conservative luminosity cuts designed to avoid the regions most sensitive to the systematics identified by K19.
The same cuts are applied to both the observed sample and the selection-function integration domain in the likelihood normalization.

\subsection{Conservative re-inclusion of the low-redshift sample}
\label{subsec:lowz}

The low-redshift interval is important for anchoring the local LF in a shape-preserving model, but it is also where host-galaxy contamination and extended-source incompleteness are most likely to affect the faint end.
We therefore do not restore the full $z<0.6$ sample.
Instead, we keep only a bright subset where the LF is less sensitive to the problematic faint-end regime.
Specifically, we impose
\begin{equation}
0.1 \le z < 0.4,\qquad M_{1450}\le -20.7,
\end{equation}
and
\begin{equation}
0.4 \le z < 0.6,\qquad M_{1450}\le -21.5.
\end{equation}
Objects outside these low-redshift luminosity cuts are excluded from the unbinned likelihood analysis.
These thresholds are chosen to retain the bright part of the corresponding low-redshift intervals while excluding the faint regimes most affected by the systematics discussed by K19.
This choice retains the local anchoring power of the low-redshift sample while avoiding the faintest part of the SDSS/2SLAQ low-redshift data, where K19 found the strongest evidence for residual systematics.

\subsection{Conservative re-inclusion of the BOSS DR9 sample}
\label{subsec:boss}

K19 excluded the BOSS DR9 quasar sample \citep{2013ApJ...773...14R} from their global evolutionary fits because of potential selection-function systematics over $2.2\le z<3.5$.
At the same time, BOSS is statistically valuable: it samples the redshift range near the peak of quasar activity and contains far more quasars than earlier samples over a similar magnitude range.
We therefore re-include BOSS only as a bright-end constraint, rather than using the full BOSS magnitude range.
To incorporate BOSS while minimizing sensitivity to uncertainties near the survey flux limit, we restrict the BOSS contribution to the bright end by applying
\begin{equation}
M_{1450} \le -23.8,
\end{equation}
and exclude objects fainter than this threshold.
This threshold is chosen to remove the region close to the BOSS selection boundary, where the inferred QLF is most sensitive to the assumed selection function.
We treat the resulting magnitude-limited BOSS subset as a separate survey component with a restricted integration domain, and incorporate its selection function consistently in both the effective-volume integral and the likelihood normalization.
The combined sample is shown in Figure~\ref{fig:sample_data}, and the before/after counts are summarized in Table~\ref{tab:data}.
By imposing this bright-end cut and consistently incorporating the corresponding selection function into the likelihood normalization, we aim to reduce sensitivity to the survey-limit systematics emphasized by K19.
We therefore treat the BOSS-inclusive analysis as a conservative extension of the K19 baseline.
Together with the low-redshift cuts in Section~\ref{subsec:lowz}, this defines the final analysis sample used throughout this work.

\section{Methodology}
\label{sec:methods}

\subsection{QLF definition and likelihood function}
\label{subsec:lf_like}

We define the QLF, $\phi(z, M)$, as the number density of quasars per unit comoving volume per unit absolute magnitude,
\begin{equation}
\phi(z, M) \equiv \frac{d^2 N}{dV dM},
\end{equation}
with units of $\mathrm{cMpc}^{-3}\mathrm{mag}^{-1}$. Hereafter, we use the shorthand notation $M \equiv M_{1450}$.
Given a model LF $\phi(z, M | \boldsymbol{\theta})$ parameterized by $\boldsymbol{\theta}$, the optimal parameter values can be obtained by maximizing the likelihood function constructed from the observed data. Following the formalism of \cite{marshall1983analysis} and \cite{2001AJ....121...54F}, the negative log-likelihood is given by
\begin{equation}
\label{eq:likelihood1}
\begin{aligned}
S =
&-2 \sum_{j}\sum_{i=1}^{n_j} \ln \left[ \phi_j(z_{i}, M_{i}) \, f_j(z_{i}, M_{i}) \right] \\
&+ 2 \sum_{j}\int\!\!\!\!\int_{W_j} \phi_j(z, M) \, f_j(z, M) \, \Omega_j \, \frac{dV}{dz} \, dz \, dM,
\end{aligned}
\end{equation}
where the index $j$ runs over survey components, $n_j$ is the number of sources in survey component $j$, $\Omega_j$ is the solid angle of component $j$, $W_j$ denotes the survey region in the $(z,M)$ space of each survey, $f_j(z,M)$ is its (homogenized) selection function, and $\mathrm{d}V/\mathrm{d}z$ is the differential comoving volume per unit solid angle defined by \cite{hogg1999distance}.
We fit the QLF parameters using Markov Chain Monte Carlo (MCMC) sampling with the Python package {\sc emcee} \citep{foreman2013emcee}, and use the resulting posterior samples to estimate the best-fitting values and credible intervals.

\begin{table*}
\tablewidth{0pt}
\renewcommand{\arraystretch}{1.5} 
\setlength{\tabcolsep}{1mm}       
\caption{Best-fit parameters for Models A and B}
\centering
\resizebox{\textwidth}{!}{%
\hspace*{-3.4cm} 
\begin{tabular}{lccccccccccc}
\hline
\hline
Model~~
& $\log_{10}(\phi_{\star})$ & $M_{\star}$
& $\beta$ & $\gamma$ & $p_1$ & $p_2$ & $p_3$ & $k_1$ & $k_2$ & $k_3$ \\
\hline
A~~
& $-6.635_{-0.029}^{+0.029}$ & $-22.345_{-0.052}^{+0.052}$
& $-1.675_{-0.013}^{+0.013}$ & $-3.819_{-0.036}^{+0.036}$
& $~~2.112_{-0.017}^{+0.017}$ & $1.222_{-0.052}^{+0.052}$
& $~~9.100_{-0.180}^{+0.180}$ & $0.066_{-0.011}^{+0.011}$
& $-0.863_{-0.012}^{+0.012}$ & $2.224_{-0.033}^{+0.033}$
\\
B~~
& $-6.586_{-0.029}^{+0.029}$ & $-21.521_{-0.041}^{+0.041}$
& $-1.588_{-0.014}^{+0.014}$ & $-3.510_{-0.027}^{+0.027}$
& $~~4.090_{-0.120}^{+0.120}$ & $-3.087_{-0.071}^{+0.071}$
& $\cdot\cdot\cdot$ & $0.179_{-0.006}^{+0.006}$
& $-1.021_{-0.007}^{+0.007}$ & $2.135_{-0.013}^{+0.013}$ \\
\hline
\hline
\end{tabular}}
\vspace{0.2cm}

\parbox{\textwidth}{%
    \raggedright\small
    Note. Units --- $\phi_{\star}$: [${\rm cMpc^{-3}~\rm{mag}^{-1}}$].
    Best-fitting parameters with $1\sigma$ uncertainties are shown
    for models A and B.%
  }
\label{modelpara}
\end{table*}

\subsection{Shape-preserving QLF models}
\label{subsec:qlf_models}

We describe the global QLF using a shape-preserving luminosity-and-density-evolution framework.
In this approach, the local LF shape is fixed, while the redshift evolution is represented by a density-evolution factor, $e_1(z)$, and a luminosity-evolution factor, $e_2(z)$.
For a luminosity-space LF defined per unit $\log_{10}L$, the model can be written as
\begin{equation}
\label{eq:shape_preserving_qlf_L}
\Phi(z,L)
=
e_1(z)\,
\Phi\left(z=0,\frac{L}{e_2(z)}\right)
\end{equation}
where $e_1(z)$ changes the overall abundance and $e_2(z)$ shifts the luminosity scale.
Equivalently, in absolute-magnitude space, this becomes
\begin{equation}
\label{eq:shape_preserving_qlf_M}
\phi(z,M)
=
e_1(z)\,
\phi\left(
z=0,
M+2.5\log_{10}\left[e_2(z)\right]
\right).
\end{equation}
Thus, $e_1(z)$ controls the vertical normalization of the LF, while $e_2(z)$ produces a horizontal shift along the magnitude axis.
This corresponds to the ``translation'' evolution form introduced by \citet{1984ApJ...284...44C}, and is closely related to the luminosity-and-density-evolution models used in previous LF studies \citep[e.g.,][]{2010MNRAS.401.2531A,yuan2017mixture,Novak_2017,2022ApJ...941...10V,2024A&A...683A.174W,2026ApJ...997..176W}.

For the local QLF shape, we adopt the conventional double power-law form widely used in UV QLF studies \citep[e.g.,][]{1988MNRAS.235..935B,2019MNRAS.488.1035K},
\begin{equation}
\label{eq:dpl_local_lf}
\phi_{\rm DPL}(z=0,M)
=
\frac{\phi_{\star}}
{
10^{0.4(\beta+1)(M-M_{\star})}
+
10^{0.4(\gamma+1)(M-M_{\star})}
},
\end{equation}
where $\phi_{\star}$ is the normalization, $M_{\star}$ is the break magnitude, and $\beta$ and $\gamma$ describe the faint- and bright-end slopes, respectively.
Under the luminosity-evolution shift in Equation~(\ref{eq:shape_preserving_qlf_M}), the break magnitude evolves as
\begin{equation}
\label{eq:mstar_evolution}
M_{\star}(z)
=
M_{\star}
-
2.5\log_{10}e_2(z).
\end{equation}

We explore a grid of candidate density- and luminosity-evolution functions for $e_1(z)$ and $e_2(z)$.
The full list of functional forms is given in Appendix~\ref{app:evolution_library}, and all functions are normalized such that $e_1(0)=1$ and $e_2(0)=1$.
This grid is used to test whether the global UV QLF can be described by a compact shape-preserving model, rather than by allowing the DPL shape parameters to vary freely with redshift.

The comparison in the main analysis is built around two DPL-based LADE models.
Model~A adopts a smooth broken-power-law density evolution,

\begin{equation}
\label{eq:modelA_density}
e_1^{\rm A}(z)
=
\mathcal{C}_{\rm A}
\frac{(1+z)^{p_2}}
{
1+\left[\dfrac{1+z}{1+p_1}\right]^{p_3}
},
\qquad
\mathcal{C}_{\rm A}
=
1+(1+p_1)^{-p_3},
\end{equation}
which has the same functional form as the parameterization commonly used to describe the cosmic star-formation history \citep{2014ARA&A..52..415M}. The luminosity-evolution form for Model~A is

\begin{equation}
\label{eq:modelA_luminosity}
e_2^{\rm A}(z)
=
\frac{
10^{-k_1k_3}+10^{-k_2k_3}
}{
10^{k_1(z-k_3)}+10^{k_2(z-k_3)}
}.
\end{equation}
Model~B keeps the same DPL local LF and the same luminosity-evolution form as Model~A, but replaces the density evolution with a smoother running-power-law form,
\begin{equation}
\label{eq:modelB_density}
e_1^{\rm B}(z)
=
(1+z)^{p_1+p_2\ln(1+z)},
\qquad
e_2^{\rm B}(z)=e_2^{\rm A}(z).
\end{equation}
Model~A is therefore the fiducial shape-preserving DPL-LADE model selected from the grid search, while Model~B serves as a density-evolution control model.
The complete model-grid rankings are reported in Table~\ref{tab:full_model_grid}.

\subsection{Flexible double-power-law reference models}
\label{subsec:fdpl_model}

As flexible reference models, we refit two redshift-dependent double-power-law (FDPL) parameterizations following K19, using the same unbinned likelihood, final analysis sample, and survey selection functions as used for Models~A and B.
Unlike the shape-preserving LADE models, the FDPL models allow the DPL normalization, break magnitude, faint-end slope, and bright-end slope to vary explicitly with redshift.
They therefore provide direct reference fits for testing whether explicit redshift evolution of the LF shape improves the global QLF description.
We consider two FDPL forms: FDPL-M1, corresponding to the preferred FDPL model of K19, and FDPL-M3, corresponding to their simpler-slope-evolution Model~3.
The use of two FDPL references allows us to test whether the comparison with the shape-preserving LADE models depends on the specific choice of flexible DPL parameterization.
\begin{equation}
\label{eq:fdpl_phi}
\phi_{\rm FDPL}(M,z)
=
\frac{\tilde{\phi}_{\star}(z)}
{
10^{0.4[\tilde{\beta}(z)+1][M-\tilde{M}_{\star}(z)]}
+
10^{0.4[\tilde{\gamma}(z)+1][M-\tilde{M}_{\star}(z)]}
},
\end{equation}
where $\tilde{\phi}_{\star}(z)$ is the redshift-dependent normalization, $\tilde{M}_{\star}(z)$ is the redshift-dependent break magnitude, and $\tilde{\beta}(z)$ and $\tilde{\gamma}(z)$ are the redshift-dependent faint- and bright-end slopes, respectively.

In the FDPL-M1 model, following K19, we describe these four redshift-dependent quantities as
\begin{equation}
\label{eq:fdpl_param_evolution}
\begin{aligned}
\log_{10}\tilde{\phi}_{\star}(z) &= F_0(1+z),\\
\tilde{M}_{\star}(z) &= F_1(1+z),\\
\tilde{\gamma}(z) &= F_2(1+z),\\
\tilde{\beta}(z) &= F_3(1+z).
\end{aligned}
\end{equation}
The first three functions are Chebyshev-polynomial expansions,
\begin{equation}
\label{eq:fdpl_cheb}
F_i(1+z)
=
\sum_{j=0}^{n_i} c_{i,j}T_j(1+z),
\qquad i=0,1,2,
\end{equation}
where $T_j$ is the Chebyshev polynomial of the first kind.
We adopt the same polynomial orders as K19: $n_0=2$ for the normalization, $n_1=3$ for the break magnitude, and $n_2=1$ for the bright-end slope.
The faint-end slope is described by a smooth broken form,
\begin{equation}
\label{eq:fdpl_faint}
F_3(1+z)
=
c_{3,0}
+
\frac{c_{3,1}}
{
10^{c_{3,3}\zeta}
+
10^{c_{3,4}\zeta}
},
\end{equation}
with
\begin{equation}
\label{eq:fdpl_zeta}
\zeta
=
\log_{10}
\left(
\frac{1+z}{1+c_{3,2}}
\right).
\end{equation}
This formulation yields a total of 14 free parameters for FDPL-M1.

In the FDPL-M3 model, following K19, we retain the quadratic, cubic, and linear Chebyshev expansions in $(1+z)$ for the evolution of $\tilde{\phi}_{\star}$, $\tilde{M}_{\star}$, and $\tilde{\gamma}$, respectively. But we simplify the evolution of the faint-end slope $\tilde{\beta}$. Specifically, $\tilde{\beta}(z)$ is assumed to be linear in $(1+z)$, which reduces the complexity of FDPL-M3 to a total of only 11 parameters.

\subsection{Model Selection}
\label{subsec:model_selection}

We compare the candidate global QLF evolution models using the Akaike and Bayesian information criteria \citep[AIC and BIC;][]{1974ITAC...19..716A,Schwarz1978}.
For each model we maximize the likelihood to obtain $\hat{\boldsymbol{\theta}}$ (Section~\ref{subsec:lf_like}) and compute
\begin{equation}
\begin{aligned}
\mathrm{AIC} &= S(\hat{\boldsymbol{\theta}})+2q,\\
\mathrm{BIC} &= S(\hat{\boldsymbol{\theta}})+q\ln N,
\end{aligned}
\end{equation}
where $S$ is defined in Equation~(\ref{eq:likelihood1}), $q$ is the number of free parameters, and $N$ is the total number of objects in the final analysis sample.
In the main comparison table, we report $\Delta\mathrm{AIC}$ and $\Delta\mathrm{BIC}$ relative to Model~A.
For the full model-grid table in Appendix~\ref{app:full_model_grid}, we report the absolute AIC and BIC values without $\Delta$ values.
A value of $\Delta\gtrsim10$ is commonly taken as strong evidence against the model with the larger criterion value \citep[e.g.,][]{burnham2002model}.
This provides a like-for-like comparison between the shape-preserving Models~A, B and the Chebyshev-based FDPL reference models, all fitted to the same sample with the same selection functions.


\begin{table*}
\centering
\begingroup
\caption{Information-criterion comparison for the representative models discussed in the main text. The differences $\Delta{\rm AIC}$ and $\Delta{\rm BIC}$ are computed relative to Model~A.}
\label{tab:main_model_comparison}
\renewcommand{\arraystretch}{1.15}

\begin{tabular*}{0.6\textwidth}{@{\extracolsep{\fill}}lccd{7.3}d{3.3}d{7.3}d{3.3}@{}}
\hline
\hline
Model & Local LF & $q$ &
\multicolumn{1}{c}{AIC} &
\multicolumn{1}{c}{$\Delta$AIC} &
\multicolumn{1}{c}{BIC} &
\multicolumn{1}{c}{$\Delta$BIC} \\
\hline

Model A
& DPL
& 10
& 2173585.729
& 0.000
& 2173677.427
& 0.000 \\

Model B
& DPL
& 9
& 2173749.379
& 163.650
& 2173831.908
& 154.481 \\

FDPL-M1
& DPL
& 14
& 2173924.480
& 338.751
& 2174052.858
& 375.431 \\

FDPL-M3
& DPL
& 11
& 2174255.037
& 669.308
& 2174355.905
& 678.478 \\
\hline
\hline
\end{tabular*}
\endgroup
\end{table*}






\begin{figure*}
\centering
\includegraphics[width=\textwidth]{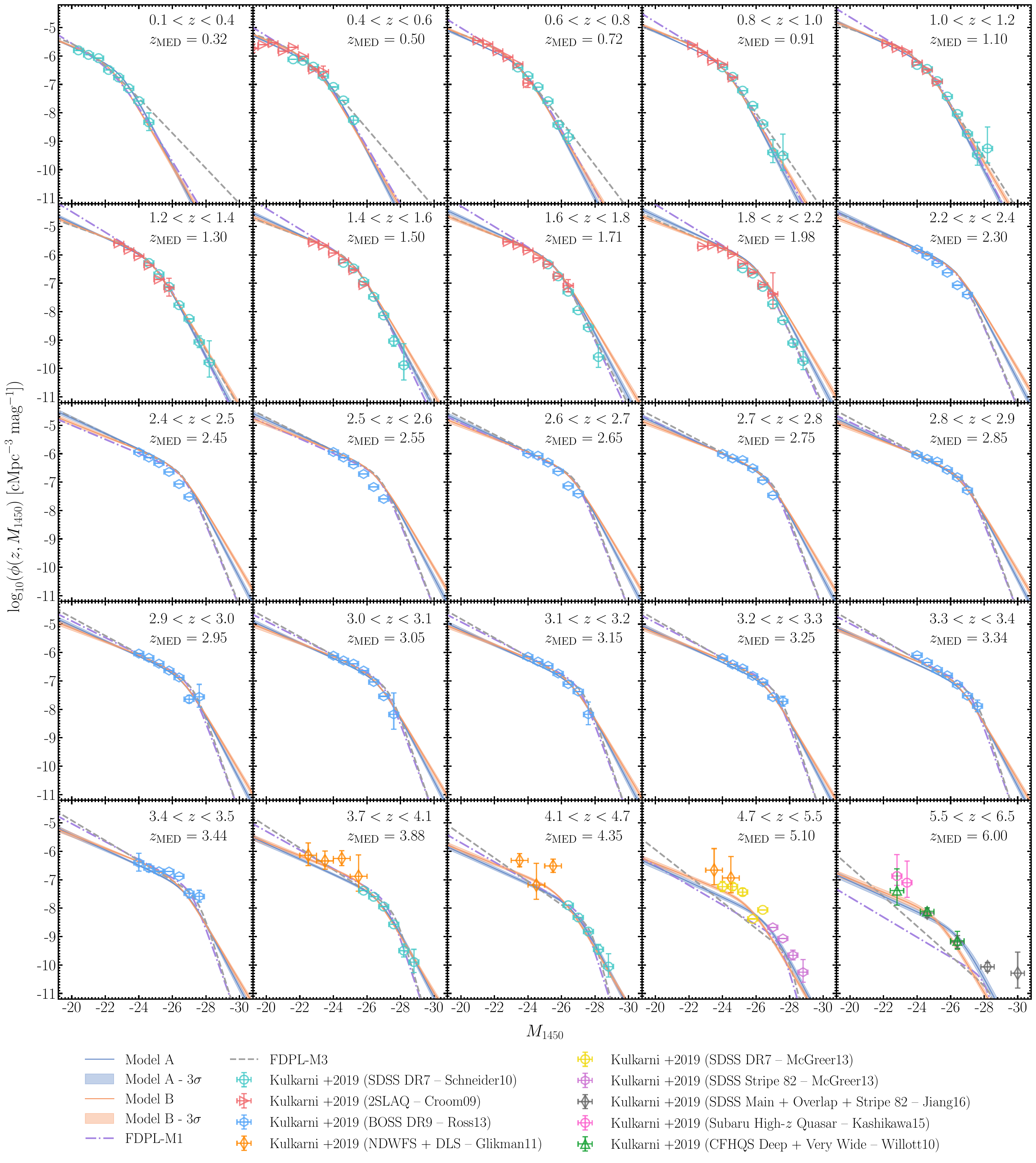}
\caption{
Best-fitting QLFs of Model~A, Model~B, and the FDPL-M1 and FDPL-M3 reference models in 25 redshift intervals.
The data points show the K19 binned QLF estimates, separated by survey component.
All model curves are evaluated at the median redshift of each bin.
The shaded regions show the $3\sigma$ credible intervals for Models~A and B.
}
\label{fig:PLF_MCMC}
\end{figure*}

\begin{figure*}
    \centering
    \gridline{
        \fig{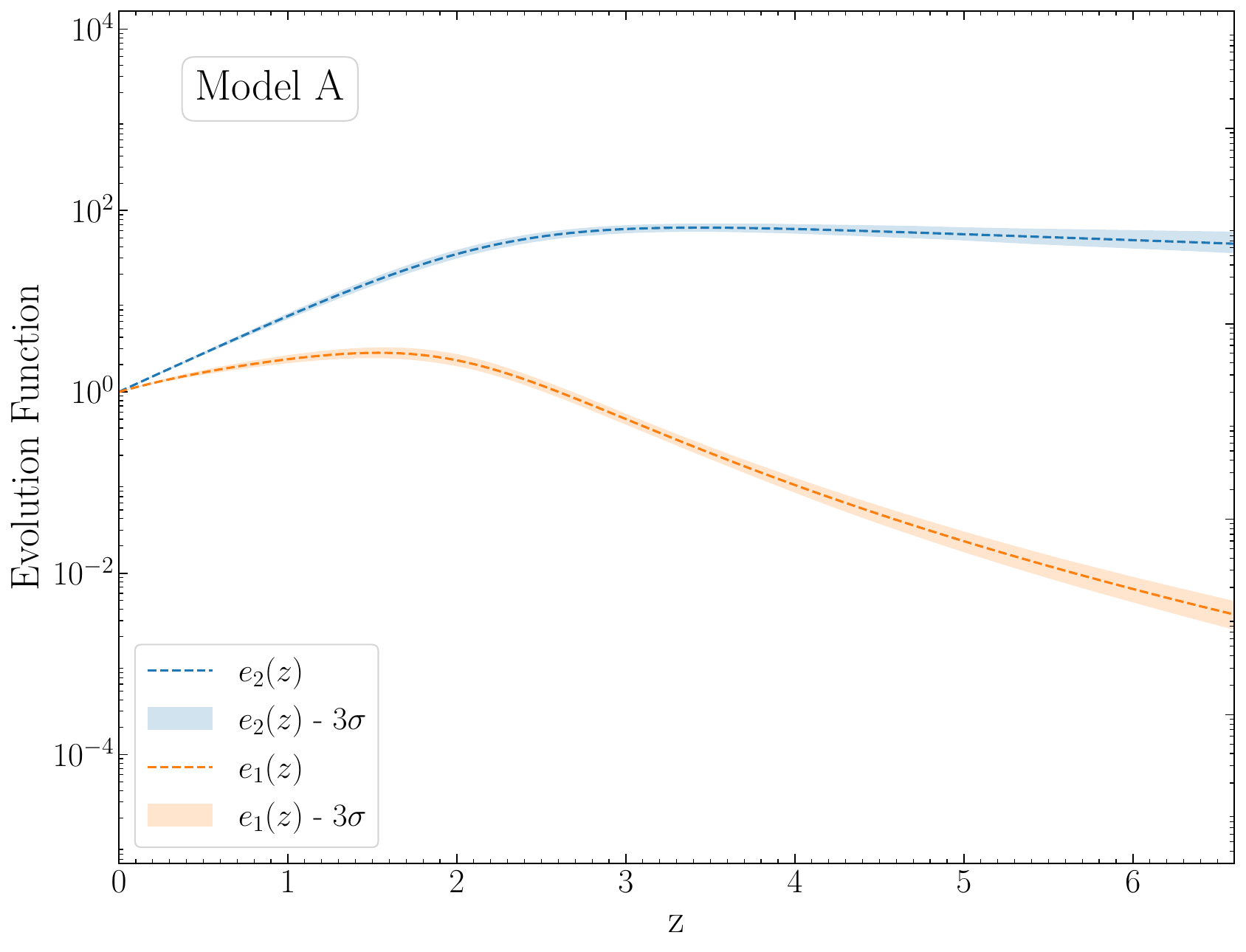}{0.48\textwidth}{}
        \hspace{-0.1in}
        \fig{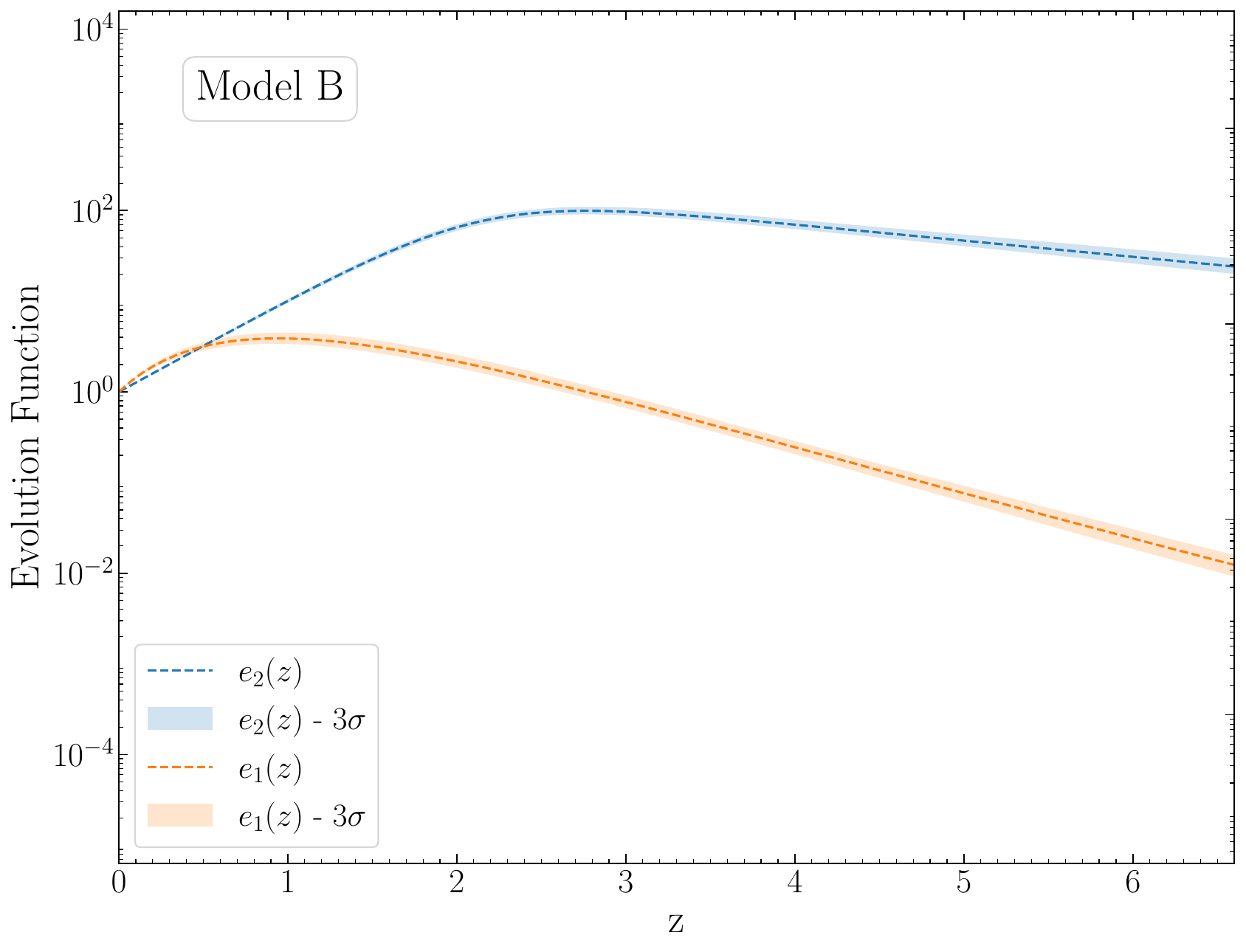}{0.48\textwidth}{}
    }
    \vspace{-0.4in} 
    \caption{
    Best-fitting density- and luminosity-evolution functions for Models A and B. The lightly shaded regions indicate the 3$\sigma$ credible intervals.}
    \label{fig:DELE}
\end{figure*}

\begin{figure}
	\centering
	\includegraphics[width=0.7\columnwidth]{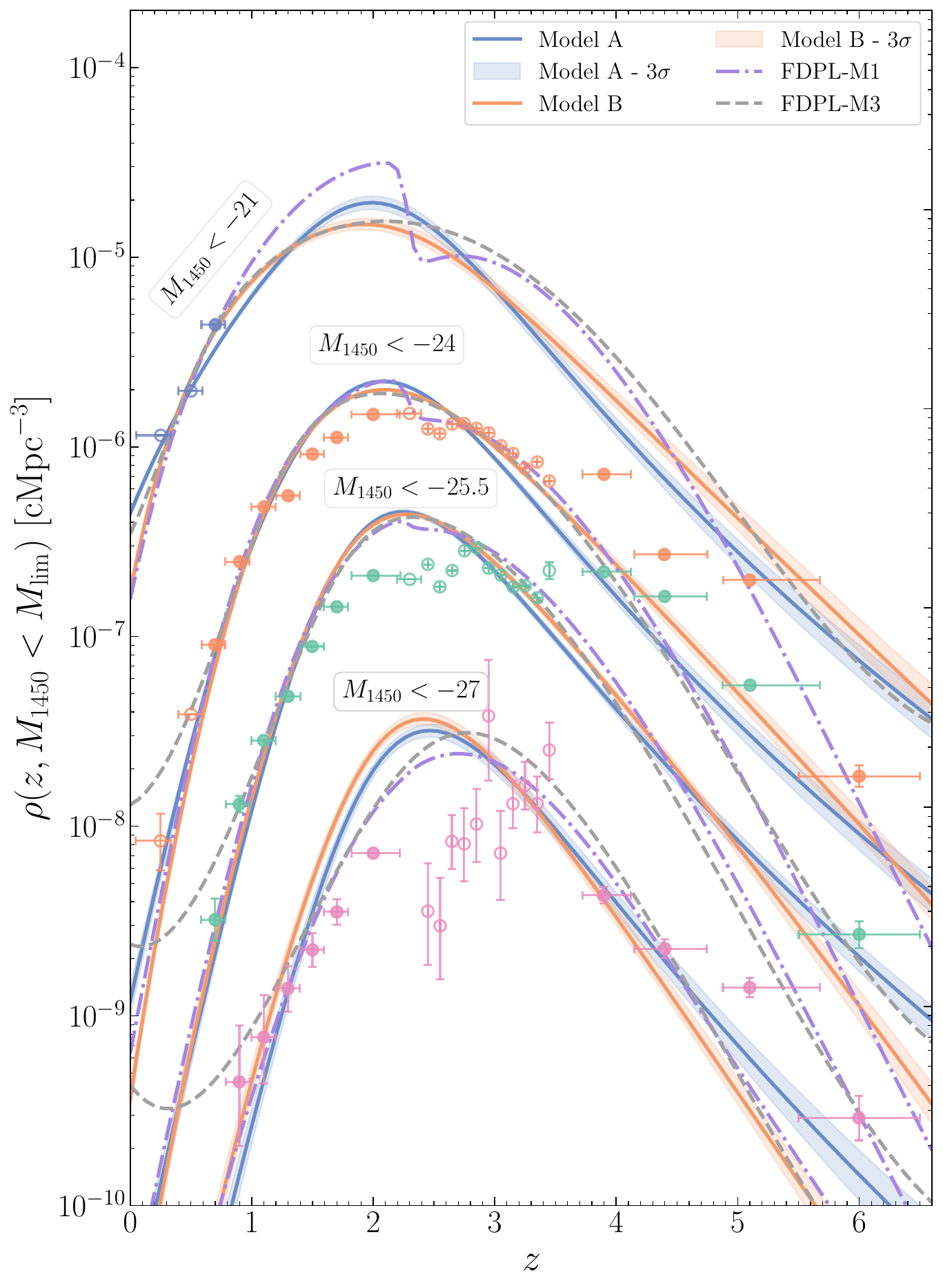}
	\hspace{-0.1in}
	\caption{
Evolution of the cumulative quasar number density brighter than several UV absolute-magnitude limits, $M_{1450}<M_{\rm lim}$, for $M_{\rm lim}=-21$, $-24$, $-25.5$, and $-27$ as labeled. Solid curves show the best-fitting predictions of Model~A (blue) and Model~B (orange), with shaded regions indicating the corresponding $3\sigma$ credible intervals. The purple dot--dashed and gray dashed curves show the FDPL-M1 and FDPL-M3 reference models, respectively, both refitted in this work following the corresponding parameterizations of K19. The data points show the binned estimates reported by K19.
    }

	\label{fig:LF_Lum}
\end{figure}

\section{Results}
\label{sec:results}

\subsection{Best-fitting QLFs}
\label{subsec:best_fit_model_comparison}

Figure~\ref{fig:PLF_MCMC} presents the best-fitting QLFs derived from Model~A (blue solid curves), Model~B (orange solid curves), and the FDPL-M1 (purple dot--dashed curves) and FDPL-M3 (gray dashed curves) reference models in 25 redshift intervals.
The model curves are evaluated at the median redshift of each bin.
The binned QLF estimates from K19, obtained using the $1/V_{\max}$ method with survey-specific selection functions and effective volumes, are shown for comparison.

The parameter constraints underlying these reconstructed QLFs are summarized by the posterior diagnostics.
For Models~A and B, the marginalized posterior distributions are shown in Figures~\ref{fig:cornerplotA} and \ref{fig:cornerplotB}, and the best-fitting parameters with $1\sigma$ credible intervals are listed in Table~\ref{modelpara}.
The corresponding posterior samples are used to construct the shaded credible regions in Figure~\ref{fig:PLF_MCMC}.
For the FDPL reference models, the posterior diagnostics and best-fitting coefficients are given in Figures~\ref{fig:cornerplotChebyshev1}--\ref{fig:cornerplotChebyshev2} and Table~\ref{modelpara_Chebyshev}.

The two shape-preserving LADE models yield remarkably similar QLF reconstructions over most of the luminosity and redshift range considered here.
The largest differences appear primarily at the faintest luminosities and at the highest redshifts, where observational constraints remain limited and the inferred QLF becomes increasingly sensitive to the adopted evolutionary form.
Within the luminosity range well sampled by existing surveys, however, the predictions of Models~A and B are nearly indistinguishable.

The FDPL reference models show more noticeable differences from the two LADE models.
In several redshift intervals, these differences appear mainly at the faint and bright ends of the QLF, where the model curves are partly extrapolated beyond the luminosity range strongly constrained by the data.
These edge differences therefore should not be over-interpreted as direct evidence for or against a particular model.
The most visible discrepancy occurs in the two highest-redshift bins, where the FDPL curves agree less well with the binned QLF estimates than the two shape-preserving LADE models.

The qualitative differences seen in Figure~\ref{fig:PLF_MCMC} are quantified by the information-criterion comparison in Table~\ref{tab:main_model_comparison}.
Model~A gives the lowest AIC and BIC values among the models shown in Figure~\ref{fig:PLF_MCMC}.
Although Model~B produces a nearly indistinguishable QLF reconstruction over most of the observed range, it is disfavored relative to Model~A by $\Delta\mathrm{AIC}=163.650$ and $\Delta\mathrm{BIC}=154.481$.
The two FDPL reference models give still larger differences.
For FDPL-M1, which corresponds to the preferred flexible DPL form of K19, we obtain $\Delta\mathrm{AIC}=338.751$ and $\Delta\mathrm{BIC}=375.431$.
For FDPL-M3, which adopts a simpler linear evolution for the faint-end slope, the differences increase to $\Delta\mathrm{AIC}=669.308$ and $\Delta\mathrm{BIC}=678.478$.
Thus, neither flexible redshift-dependent DPL reference model improves the global fit once model complexity is taken into account.

Taken together, the QLF reconstruction and the information-criterion comparison show that the observed evolution of the UV QLF can be reproduced without introducing explicit redshift evolution in the intrinsic LF shape.
Within the model library and likelihood framework adopted here, a shape-preserving local LF combined with separate density and luminosity evolution factors provides a statistically preferred and interpretable description of the current quasar sample over the redshift range $0.1 \lesssim z \lesssim 7.5$.

\subsection{Density--luminosity decomposition}
\label{subsec:evolution_functions}

Figure~\ref{fig:DELE} shows the fitted density- and luminosity-evolution functions for Models~A and B.
In the shape-preserving LADE framework, $e_1(z)$ represents the density-evolution factor, while $e_2(z)$ represents the luminosity-evolution factor.
The luminosity evolution is very similar in the two models: $e_2(z)$ rises rapidly from the local universe to $z\simeq2$--3 and then becomes nearly flat or slowly declines at higher redshift.
This common behavior is consistent with the close agreement between the reconstructed QLFs of Models~A and B shown in Figure~\ref{fig:PLF_MCMC}.

The density-evolution factor shows a stronger dependence on the adopted functional form.
In Model~A, $e_1(z)$ varies only mildly at low redshift and then declines toward higher redshift.
In Model~B, $e_1(z)$ rises more noticeably at low redshift before turning over and decreasing at higher redshift.
Thus, while the luminosity-evolution history is nearly the same in the two models, the density-evolution history is more model dependent.

This density--luminosity decomposition should be interpreted phenomenologically.
The two functions $e_1(z)$ and $e_2(z)$ are inferred jointly from the full QLF fit and are not independently measured observables.
A luminosity shift can partly mimic a change in number density over a limited luminosity range, so Figure~\ref{fig:DELE} should be read as an empirical decomposition of the fitted global QLF evolution.

\subsection{Cumulative Number Density Evolution}
\label{subsec:results_cumdens}

Figure~\ref{fig:LF_Lum} shows the evolution of the cumulative quasar number density obtained by integrating the QLF down to $M_{\rm lim}=-21$, $-24$, $-25.5$, and $-27$,
\begin{equation}
\label{eq:cumdens}
\rho\!\left(z,\, M_{1450}<M_{\rm lim}\right)
= \int_{-\infty}^{M_{\rm lim}}
\phi\!\left(M_{1450},z\right)\, dM_{1450}.
\end{equation}
The curves show the predictions of Models~A and B and of the FDPL reference models, with shaded regions denoting the $3\sigma$ credible intervals. The points show the binned number-density estimates from K19, with open symbols denoting bins affected by systematics.

Across all magnitude limits, the cumulative number density rises rapidly from low redshift, reaches a broad maximum at $z\simeq2$--3, and then declines toward higher redshift.
As expected, the overall normalization decreases for brighter magnitude limits.
Models~A and B give very similar cumulative histories over the full range of $M_{\rm lim}$, indicating that the integrated number-density evolution is stable between the two shape-preserving LADE models.

The FDPL reference models show more pronounced differences from Models~A and B, and the form of the difference varies with the adopted magnitude limit.
For the fainter limits, especially $M_{\rm lim}=-21$ and $-24$, the FDPL-M1 curve exhibits a localized, non-smooth feature around the peak epoch at $z\simeq2$--3, followed by a relatively rapid decline toward higher redshift.
Such a sharp feature is difficult to interpret as a robust physical signature of quasar evolution, and more likely reflects the additional freedom of the FDPL parameterization, in which the DPL slopes are allowed to vary explicitly with redshift.
This interpretation is also consistent with the direct QLF comparison in Figure~\ref{fig:PLF_MCMC}, where the FDPL models change the relative contribution of the faint and bright ends to the integrated number density.

The K19 binned number-density estimates broadly follow the same rise-and-decline evolution as the smooth model predictions but show larger bin-to-bin fluctuations, especially around the peak epoch at $z\simeq2$--3 and at the highest redshifts.
Models~A and B provide a smoother cumulative history and tend to lie above the K19 estimates near the peak for the intermediate magnitude limits, whereas some high-redshift K19 estimates lie above the smooth model curves.
These differences likely reflect the binning and selection-function systematics inherited from the K19 binned QLF estimates, rather than a simple discrepancy in the global evolutionary trend.

\section{Discussion}
\label{sec:discussion}

\subsection{A shape-preserving empirical description of the global UV QLF}
\label{subsec:discussion_shape_preserving}

The main result of this work is that the global UV QLF can be described without requiring explicit redshift evolution in the intrinsic DPL shape.
In our fiducial Model~A, the local LF is represented by a DPL whose shape parameters remain fixed with redshift, while the cosmic evolution is captured by a density-evolution factor, $e_1(z)$, and a luminosity-evolution factor, $e_2(z)$.
This model gives the lowest AIC and BIC among the representative models compared in Table \ref{tab:main_model_comparison}, including the FDPL reference fits in which the normalization, break magnitude, faint-end slope, and bright-end slope are all allowed to vary explicitly with redshift.

This comparison is non-trivial because the FDPL models have greater formal flexibility.
They can reproduce redshift-dependent changes in both the faint and bright ends of the QLF, whereas Model~A restricts the evolution to a luminosity shift and an effective normalization change.
The fact that Model~A remains preferred after accounting for model complexity indicates that the present data do not require explicit redshift evolution in the DPL slopes.

This result should be interpreted as an empirical statement rather than as proof of a strictly invariant physical LF shape.
In the factorized LADE construction, the local LF sets the baseline shape, $e_2(z)$ shifts the characteristic luminosity scale, and $e_1(z)$ describes the residual normalization evolution after this luminosity shift is applied.
These components are inferred jointly from the full QLF fit and should not be regarded as independently measured physical quantities.
Thus, the success of Model~A shows that a common local LF shape, combined with separate density and luminosity evolution factors, provides an economical description of the observed QLF evolution over the selected $M_{1450}$--$z$ range.
A related question is whether this conclusion is tied specifically to the DPL local LF form, which we examine below.

\subsection{Local LF form and evolutionary robustness}
\label{subsec:discussion_local_lf}

The factorized LADE framework separates the baseline LF shape from the redshift-dependent evolution factors.
This makes it possible to test whether the preferred evolutionary structure is driven by the assumed local LF form or by the global distribution of quasars in the $M_{1450}$--$z$ plane.
To address this point, Appendix~\ref{app:full_model_grid} repeats the full model-grid analysis using the Saunders modified-Schechter local LF, based on the functional form introduced by \citet{saunders199060}.
The mathematical form of this alternative local LF and the full model rankings are given in the appendix.

\begin{figure*}
    \centering
    \gridline{
        \fig{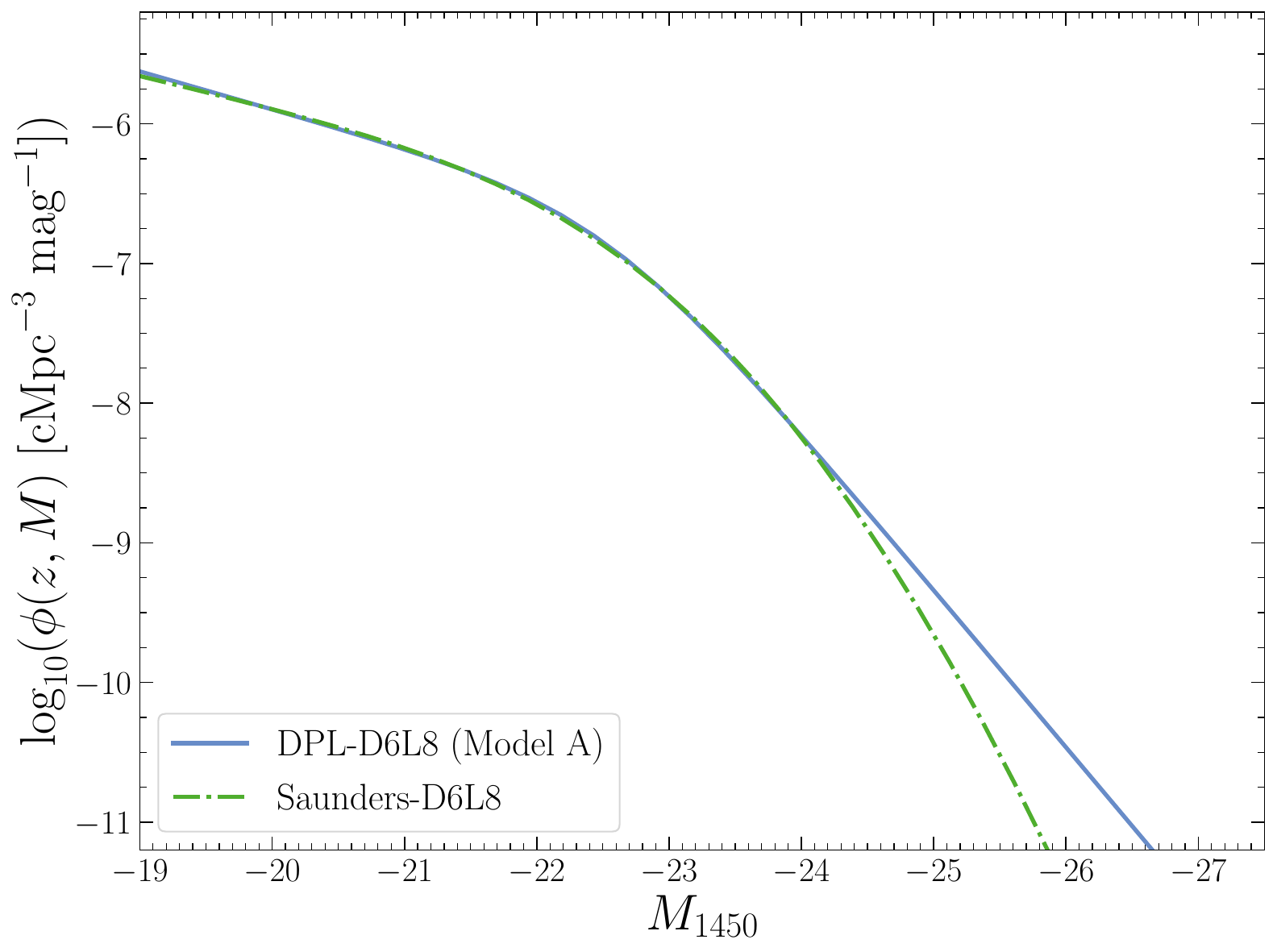}{0.48\textwidth}{}
        \hspace{-0.1in}
        \fig{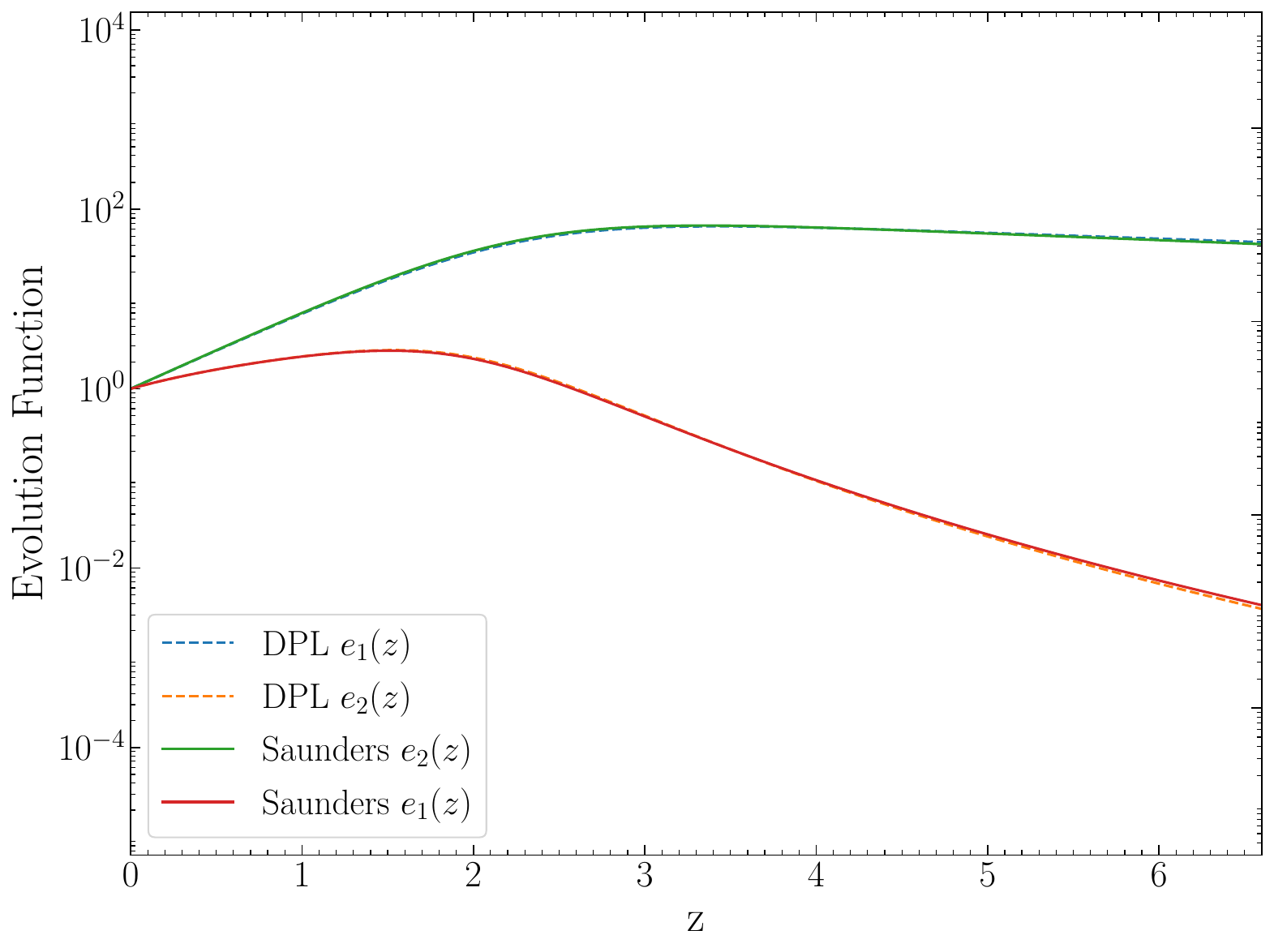}{0.48\textwidth}{}
    }
    \vspace{-0.4in} 
    \caption{Comparison between the DPL-D6L8 and Saunders-D6L8 fits.
Left: best-fitting QLFs evaluated at $z=0.1$.
Right: corresponding density- and luminosity-evolution functions.
     }
    \label{fig:comp_DPL_Sau}
\end{figure*}

This comparison shows that the local LF form affects the absolute quality of the fit, but not the preferred evolutionary structure.
The DPL local LF gives the lowest AIC and BIC values overall and is therefore adopted as our fiducial baseline shape.
However, the best-ranked model in the Saunders branch also selects the same density--luminosity evolution combination as Model~A, which is labeled D6L8 in the model-grid notation introduced in Appendix~\ref{app:evolution_library}.
This behavior is illustrated in Figure~\ref{fig:comp_DPL_Sau}.
The left panel shows that the two local LF choices are not identical: although the DPL-D6L8 and Saunders-D6L8 QLFs are similar over the intermediate luminosity range, they differ clearly at the bright end, where the Saunders local LF has a steeper cutoff than the DPL form.
Despite this difference in the baseline LF shape, the right panel shows that the corresponding $e_1(z)$ and $e_2(z)$ curves are nearly indistinguishable between the two local LF choices.
Thus, changing the local LF form mainly affects the detailed LF shape and the absolute goodness of fit, while the inferred global density- and luminosity-evolution histories remain stable within the two local-LF families tested here.

\begin{figure*}
    \centering
    \gridline{
        \fig{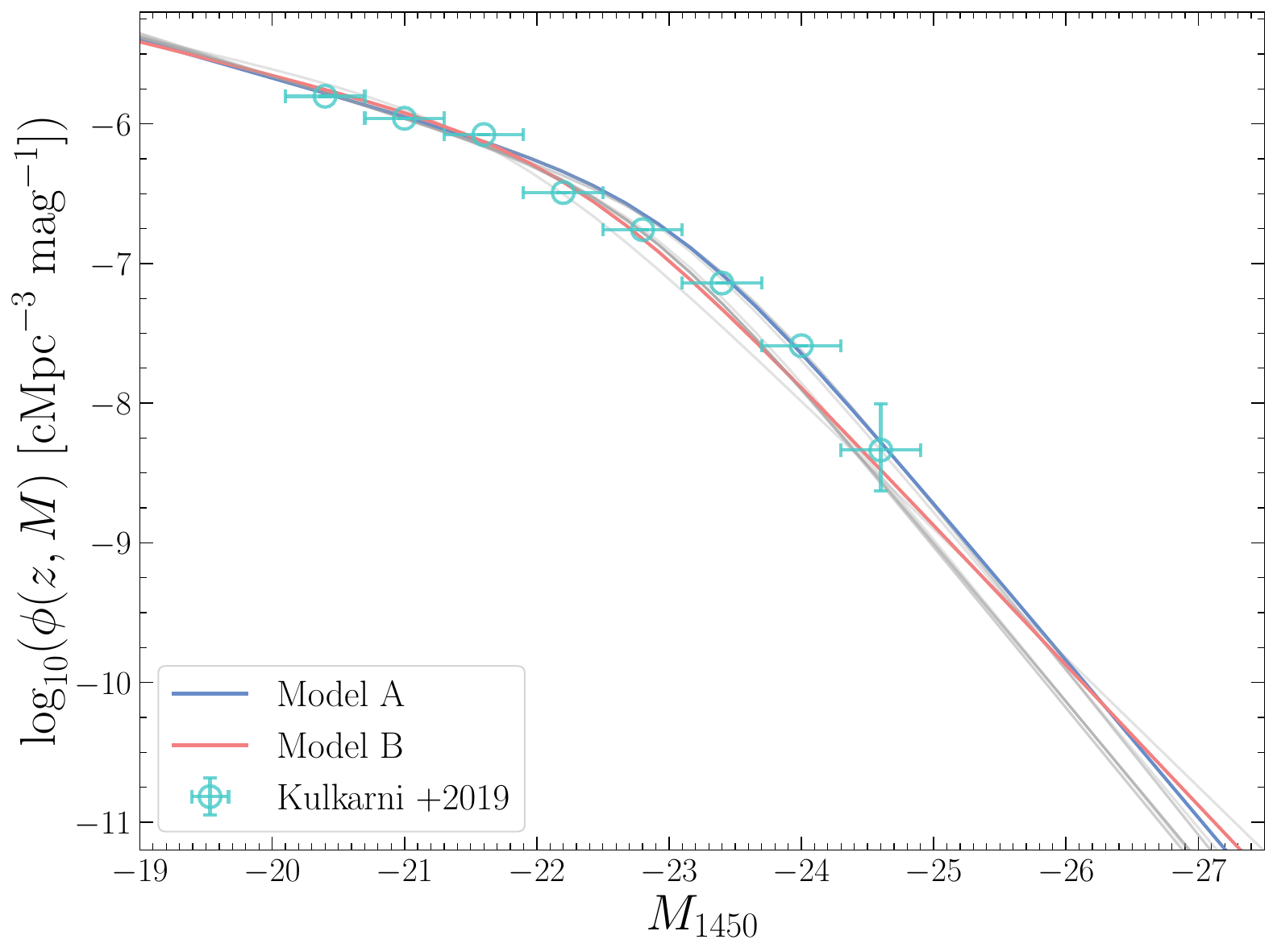}{0.48\textwidth}{}
        \hspace{-0.1in}
        \fig{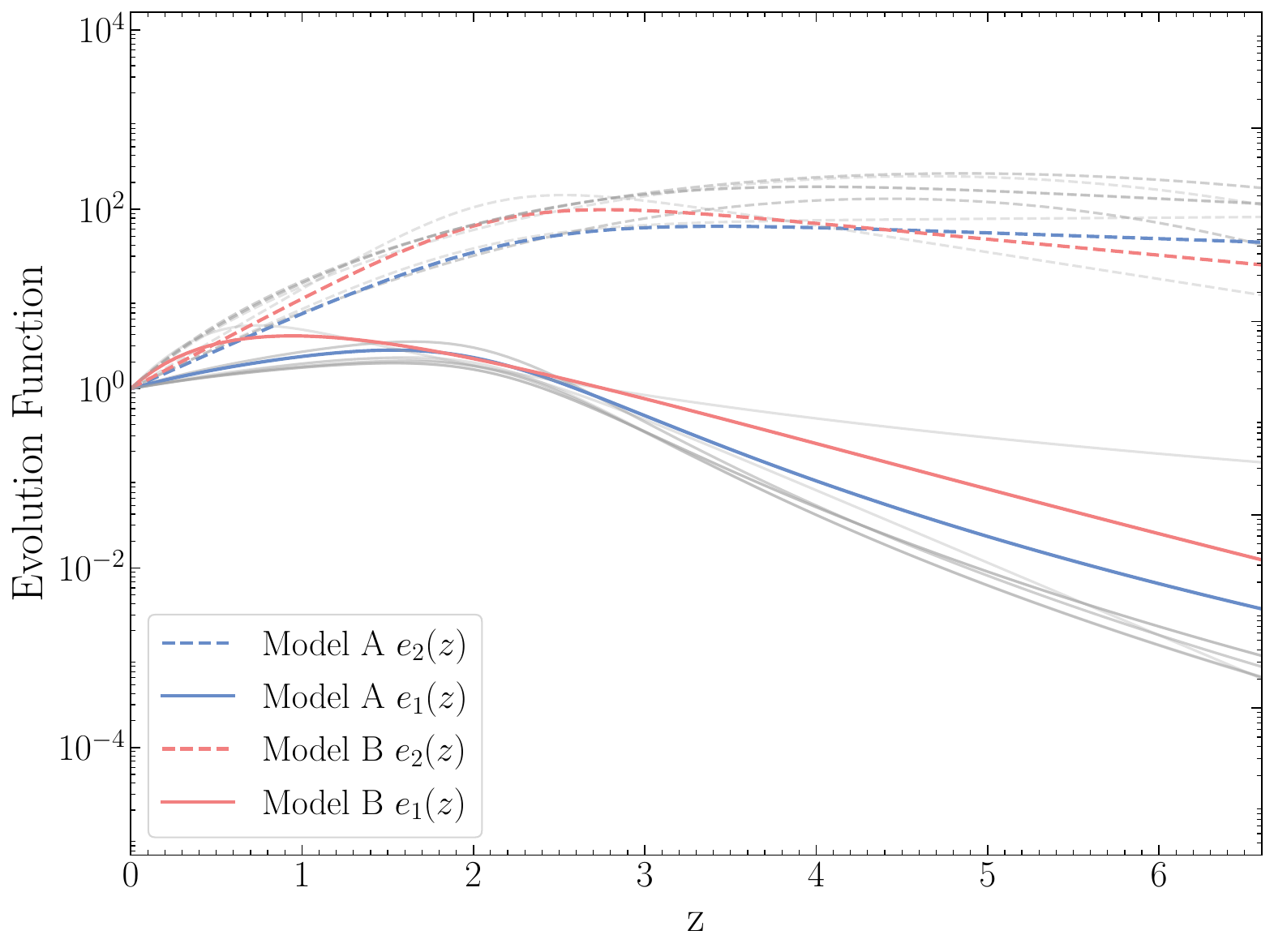}{0.48\textwidth}{}
    }
    \vspace{-0.4in} 
    \caption{Robustness of the DPL-based LADE model family.
    The grey curves show the 12 DPL-based LADE models whose AIC and BIC values are lower than those of the refitted FDPL-M1 reference model; Models~A and B are included in this set and are highlighted in color. Left: best-fitting QLFs evaluated at the median redshift of the first bin, $z_{\rm MED}=0.32$, together with the binned estimates reported by K19.
Right: corresponding density- and luminosity-evolution functions, where solid curves denote $e_1(z)$ and dashed curves denote $e_2(z)$. The reconstructed QLFs and luminosity-evolution histories are broadly stable across this model family, while the density-evolution component is more sensitive to the adopted functional form.
    }
    \label{fig:comp_LLF_DELE}
\end{figure*}

\subsection{Density--luminosity decomposition and model dependence}
\label{subsec:discussion_de_le}

The density--luminosity decomposition in Section~\ref{subsec:evolution_functions} shows that Models~A and B have very similar luminosity-evolution histories but different density-evolution histories.
Here we further examine whether this behavior is specific to these two representative models or is a broader property of the preferred DPL-based LADE model family.

Figure~\ref{fig:comp_LLF_DELE} compares the 12 DPL-based LADE models whose AIC and BIC values are lower than those of the FDPL reference models.
Models~A and B are included in this set and are highlighted in color, while the other models are shown in grey.
The left panel shows that the reconstructed QLFs at the median redshift of the first bin, $z_{\rm MED}=0.32$, are very similar over the luminosity range directly constrained by the binned estimates.
The largest differences appear mainly toward the bright end, where the constraints are weaker and the result is more sensitive to extrapolation.
The right panel shows the corresponding evolution functions.
The luminosity-evolution curves, $e_2(z)$, have broadly similar shapes across this model family, whereas the density-evolution curves, $e_1(z)$, show larger model-to-model variations, especially at high redshift.

This comparison indicates that the preference for a shape-preserving LADE description is not driven by a single best-fitting model.
Within the DPL-based LADE models that outperform both of the FDPL reference models, the reconstructed QLFs and luminosity-evolution histories are relatively stable, while the detailed density-evolution history remains more sensitive to the adopted functional form.
This reinforces the interpretation of the LADE factors as an economical empirical decomposition of the global UV QLF evolution, rather than as a unique physical separation between luminosity growth and abundance evolution.

Similar density--luminosity decompositions have also been used in radio AGN LF studies by \citet{yuan2016mixture,yuan2017mixture}, suggesting that such LADE-type descriptions may provide a useful empirical language for comparing different AGN populations.
However, the physical interpretation of the two evolution factors is necessarily population- and wavelength-dependent.

\subsection{Low-redshift restricted-likelihood diagnostic}
\label{subsec:lowz_restricted_likelihood}
As a further check, we evaluated the restricted contribution to the negative log-likelihood in Equation~\eqref{eq:likelihood1} over the $z<2.2$ subset.
We choose this redshift range because it lies below the BOSS DR9 interval and is dominated by the low- and intermediate-redshift survey components, primarily SDSS DR7 and 2SLAQ, where the sample is relatively well populated and the selection functions are better controlled.
This test therefore provides a useful diagnostic of whether the shape-preserving result remains competitive in the relatively well-sampled low-redshift regime, independently of the BOSS and high-redshift parts of the compilation.

We denote the restricted likelihood contribution as $S_{z<2.2}$ and compute it using the best-fitting parameters obtained from the full-sample fits.
This diagnostic is not used as an independent model-selection criterion, because the full redshift-dependent evolution parameters are not re-estimated from the restricted data range.
We find that the low-redshift subset favors relatively simple and smooth evolutionary forms.
Although FDPL-M3 gives a smaller $S_{z<2.2}$ than the globally preferred Models~A and B, 31 of the 81 DPL-based LADE models defined by the model library in Appendix~\ref{app:evolution_library} yield lower $S_{z<2.2}$ values than FDPL-M3 in this diagnostic.
In contrast, FDPL-M1 is less successful than nearly all DPL-based LADE models over the same redshift range.
Thus, the low-redshift data do not by themselves require explicit LF-shape evolution.
Rather, they support the competitiveness of the shape-preserving LADE family, while the specific choice of Models~A and B is determined by the requirement to describe the full $0.1\le z\le7.5$ redshift baseline.

\subsection{Limitations of the present analysis}
\label{subsec:discussion_limitations}

Several limitations should be kept in mind when interpreting our results.
First, our data selection differs from the fiducial global-fit choice of K19.
We re-include the low-redshift interval $0.1\le z<0.6$ and a bright-end subset of the BOSS DR9 sample because these regimes provide useful leverage on the local LF anchor and on the epoch of peak quasar activity.
However, both are included conservatively, with luminosity cuts designed to reduce sensitivity to the faint low-redshift regime and to the BOSS selection boundary where K19 identified significant systematics.

Second, the comparison with the K19 binned number-density estimates should not be interpreted as an additional fit to those points.
Our model curves are derived from a single smooth unbinned likelihood fit to the selected sample, whereas the K19 estimates inherit the binning, luminosity coverage, and effective-volume corrections of the underlying binned QLF measurements.
Differences near the peak of the cumulative number-density evolution and at the highest redshifts therefore reflect, at least in part, the different ways in which the data and selection functions enter the two reconstructions.

Third, the high-redshift QLF remains weakly constrained at the faint end.
At $z\gtrsim5$, the available samples contain fewer objects and span a narrower luminosity baseline than at low and intermediate redshifts.
Consequently, cumulative densities evaluated at relatively faint limits should be regarded as more model dependent than the evolution over the well-sampled luminosity range.

Finally, the preference for Model~A is conditional on the model library, data selection, and likelihood framework adopted here.
The LADE factors $e_1(z)$ and $e_2(z)$ provide an economical empirical decomposition of the global QLF evolution, but they should not be interpreted as a unique physical separation between abundance evolution and luminosity evolution.
Future surveys with better-controlled selection functions and deeper high-redshift coverage will be needed to test whether explicit LF-shape evolution is required beyond the present shape-preserving description.

\section{Conclusions}
\label{sec:conclusions}

We have presented a global unbinned-likelihood analysis of the rest-frame ultraviolet (UV) quasar luminosity function (QLF) using the K19 homogenized Type~1 quasar compilation and survey-specific selection functions.
Relative to the fiducial K19 global-fit configuration, we conservatively re-include the low-redshift interval $0.1\le z<0.6$ and a bright-end subset of the BOSS DR9 sample, yielding a final analysis sample of 70,960 Type 1 quasars.
We model the QLF with a shape-preserving luminosity and density evolution (LADE) framework, in which the local luminosity function (LF) shape is fixed and the redshift evolution is described by separate density- and luminosity-evolution factors.
For the double-power-law (DPL) local LF, we construct a model library from nine candidate density-evolution functions and nine candidate luminosity-evolution functions, giving 81 LADE combinations in total.
The preferred DPL-based LADE models are compared with the K19-style flexible double-power-law (FDPL) reference models refitted in this work using the same sample, likelihood formalism, and survey selection functions.

Our main conclusions are as follows.

\begin{enumerate}
\item
The global UV QLF can be described by a compact shape-preserving LADE model.
The fiducial Model~A gives the lowest Akaike information criterion (AIC) and Bayesian information criterion  (BIC) values among the main models considered in the text, outperforming both the density-evolution control model and the FDPL reference models.
More broadly, 12 DPL-based LADE models have lower AIC and BIC values than the better-performing FDPL-M1 reference model, and hence also than FDPL-M3, indicating that the preference for a shape-preserving description is not confined to a single best-fitting parameterization. A restricted-likelihood diagnostic over the relatively well-sampled $z<2.2$ subset further shows that the shape-preserving LADE family remains competitive independently of the BOSS and high-redshift parts of the compilation.
Thus, within the adopted data selection, likelihood framework, and model library, the present data do not require explicit redshift evolution of the DPL faint- and bright-end slopes.

\item
The inferred evolutionary structure is not driven solely by the assumed local LF form.
Although the DPL local LF gives the best absolute fit and is adopted as the fiducial baseline shape, repeating the model-grid analysis with the Saunders modified-Schechter local LF leads to the same preferred density--luminosity evolution structure.
The corresponding $e_1(z)$ and $e_2(z)$ curves are nearly indistinguishable between the two local LF choices, showing that the main evolutionary trend is stable against this change in the baseline LF shape.

\item
The luminosity-evolution history is the most stable part of the LADE decomposition.
Across Models~A and B, and more generally across the preferred DPL-based LADE model family, $e_2(z)$ rises rapidly from the local universe to $z\simeq2$--3 and then becomes nearly flat or slowly declines.
The detailed form of the density-evolution component is more sensitive to the adopted functional form, especially at low to intermediate redshift where luminosity shifts and normalization changes are partially degenerate.
Nevertheless, the preferred LADE models consistently require the effective normalization to decline toward high redshift after the luminosity evolution has been accounted for.
This high-redshift decline is consistent with the observed decrease in quasar abundance beyond the peak epoch.

\item
The highest-redshift and faint-end regimes remain weakly constrained.
At $z\gtrsim5$--6, the data contain fewer quasars and cover a narrower luminosity range, making faint-end extrapolations more model dependent. Thus, the shape-preserving framework provides a statistically preferred global description for the current data, while its validity in the weakly constrained high-redshift faint-end regime will require deeper future samples.

\end{enumerate}

Overall, our results support a simple and statistically efficient description of the global UV QLF: a DPL local LF whose evolution is captured primarily by luminosity and density shifts.
Future larger and more homogeneous samples, especially at the high-redshift faint end, will be essential for testing whether this shape-preserving description remains valid beyond the luminosity and redshift range currently well constrained.


\begin{acknowledgments}
We acknowledge financial support from the Science Fund for Distinguished Young Scholars of Hunan Province (Grant No. 2024JJ2040), and the Major Basic Research Project of Hunan Province (Grant No. 2024JC0001). Z.Y. is supported by the Xiaoxiang Scholars Programme of Hunan Normal University. We thank the authors of \citet{2019MNRAS.488.1035K} for publicly releasing the homogenized quasar compilation and associated selection functions used in this work.
\end{acknowledgments}

\FloatBarrier  

\bibliography{refs}{}
\bibliographystyle{aasjournal}

\FloatBarrier  
\appendix

\renewcommand{\thetable}{\thesection.\arabic{table}}
\renewcommand{\thefigure}{\thesection.\arabic{figure}}
\renewcommand{\theequation}{\thesection.\arabic{equation}}

This appendix provides supplementary material for the model definitions and fitting results presented in the main text.
Appendix~\ref{app:evolution_library} lists the candidate density-evolution and luminosity-evolution functions used in the luminosity and density evolution (LADE) model-grid search, summarized in Table~\ref{tab:evolution_function_library}.
Appendix~\ref{app:Saunders} defines the alternative Saunders modified-Schechter local luminosity function (LF).
Appendix~\ref{app:full_model_grid} reports the full Akaike information criterion (AIC) and Bayesian information criterion  (BIC) rankings for the double-power-law (DPL) and Saunders-like local LF branches.
Appendix~\ref{app:posterior_diagnostics} presents the posterior diagnostics and best-fitting parameter tables for Models~A--B and for the Chebyshev-polynomial flexible double-power-law (FDPL) reference models.

\section{Candidate density- and luminosity-evolution functions}
\label{app:evolution_library}
\setcounter{table}{0}
\setcounter{figure}{0}
\setcounter{equation}{0}

In the model-grid search, we consider nine candidate density-evolution functions and nine candidate luminosity-evolution functions, denoted as D1--D9 and L1--L9, respectively.
For a given choice of the local LF form, each density-evolution function is combined with each luminosity-evolution function through Eq.~\ref{eq:shape_preserving_qlf_M} to construct a shape-preserving LADE model.
Thus, each local LF form gives $9\times9=81$ candidate evolutionary models.
For example, the label D6L8 denotes the model obtained by combining density evolution D6 with luminosity evolution L8.
The fiducial Model~A discussed in the main text corresponds to D6L8 with a DPL local LF, while Model~B corresponds to D5L8 with the same local LF form.
All functions are normalized such that $e_1(0)=1$ and $e_2(0)=1$, so that the fitted local LF parameters retain their meaning at $z=0$.

\begin{table*}[h]
\centering
\caption{
Candidate density- and luminosity-evolution functions used in the LADE model-grid search.}
\label{tab:evolution_function_library}
\scriptsize
\hspace*{-1.3cm}%
\renewcommand{\arraystretch}{1.35}
\resizebox{\textwidth}{!}{%
\begin{tabular}{cl@{\hspace{1.0em}}!{\color{gray!25}\vrule width 0.5pt}@{\hspace{1.0em}}cl}
\hline
\hline
\noalign{\vskip 0.35em}
\shortstack{\normalsize Density\\ \normalsize label}
& \raisebox{1.5ex}{\normalsize Density-evolution function}
& \shortstack{\normalsize Luminosity\\ \normalsize label}
& \raisebox{1.5ex}{\normalsize Luminosity-evolution function} \\
\noalign{\vskip 0.35em}
\hline

\normalsize D1 & \mcellbig[1.12]{e^{p_1 z}}
& \normalsize L1 & \mcellbig[1.12]{(1+z)^{k_1}} \\[0.5em]

\normalsize D2 & \mcellbig[1.12]{(1+z)^{p_1}}
& \normalsize L2 & \mcellbig[1.12]{(1+z)^{k_1+k_2 z}} \\[0.5em]

\normalsize D3 & \mcellbig[1.12]{e^{p_1 z+p_2 z^2}}
& \normalsize L3 & \mcellbig[1.12]{(1+z)^{k_1+k_2\ln(1+z)}} \\[0.5em]

\normalsize D4 & \mcellbig[1.12]{(1+z)^{p_1+p_2 z}}
& \normalsize L4 & \mcellbig[1.12]{10^{k_1 z+k_2 z^2}} \\[0.5em]

\normalsize D5 &
\mcellbig[1.12]{(1+z)^{p_1+p_2\ln(1+z)}}
& \normalsize L5 &
\mcellbig[1.12]{(1+z)^{k_1}\exp\!\left(-\dfrac{z^2}{2k_2^2}\right)} \\[1.5em]

\normalsize D6 &
\mcellbig[1.12]{
\dfrac{\left[1+(1+p_1)^{p_3}\right](1+z)^{p_2}}
{(1+p_1)^{p_3}+(1+z)^{p_3}}
}
& \normalsize L6 &
\mcellbig[1.12]{
\dfrac{\left[1+(1+k_1)^{k_3}\right](1+z)^{k_2}}
{(1+k_1)^{k_3}+(1+z)^{k_3}}
} \\[1.8em]

\normalsize D7 &
\mcellbig[1.12]{
(1+z)^{p_1}\exp\!\left(-\dfrac{z^2}{2p_2^2}\right)
}
& \normalsize L7 &
\mcellbig[1.12]{
(1+k_1)^{k_2}
\left(\dfrac{1+z}{1+k_1}\right)^{k_2}
\exp\!\left[-\left(\dfrac{z}{k_1}\right)^{k_3}\right]
} \\[1.8em]

\normalsize D8 &
\mcellbig[1.15]{
\dfrac{10^{-p_1p_3}+10^{-p_2p_3}}
{10^{p_1(z-p_3)}+10^{p_2(z-p_3)}}
}
& \normalsize L8 &
\mcellbig[1.15]{
\dfrac{10^{-k_1k_3}+10^{-k_2k_3}}
{10^{k_1(z-k_3)}+10^{k_2(z-k_3)}}
} \\[1.8em]

\normalsize D9 &
\mcellbig[1.12]{
\left[(1+p_1)^{p_2}+(1+p_1)^{p_3}\right]
\left[
\left(\dfrac{1+p_1}{1+z}\right)^{p_2}
+
\left(\dfrac{1+p_1}{1+z}\right)^{p_3}
\right]^{-1}
}
& \normalsize L9 &
\mcellbig[1.12]{
\left[(1+k_1)^{k_2}+(1+k_1)^{k_3}\right]
\left[
\left(\dfrac{1+k_1}{1+z}\right)^{k_2}
+
\left(\dfrac{1+k_1}{1+z}\right)^{k_3}
\right]^{-1}
} \\[1.7em]

\hline
\hline
\end{tabular}%
}
\end{table*}

\section{Alternative local LF: the Saunders modified-Schechter form}
\label{app:Saunders}

As an alternative to the conventional DPL local shape, we also consider the Saunders modified-Schechter local LF, following the functional form introduced by \citet{saunders199060}.
This form, often referred to as the Saunders function, has been widely used to describe the local LFs of star-forming galaxies
\citep[e.g.,][]{Novak_2017,Ocran2020,enia2022new,2022ApJ...941...10V,2023MNRAS.523.6082C,2024A&A...683A.174W,2026ApJ...997..176W}.
It has also recently been applied to the AGN radio LFs \citep[e.g.,][]{2024A&A...684A..19S,2026arXiv260315449W}.
In luminosity space, the local LF is written as
\begin{eqnarray}
\label{eq:saunders_lum}
\begin{aligned}
\Phi_{\rm Sau}(z=0,L)
=
\frac{dN}{d\log_{10}L}
=
\Phi_{\star}
\left(\frac{L}{L_{\star}}\right)^{1-\beta}
\exp\left[
-\frac{1}{2\gamma^2}
\log_{10}^{2}
\left(
1+\frac{L}{L_{\star}}
\right)
\right],
\end{aligned}
\end{eqnarray}

where $L_{\star}$ marks the knee of the LF, $\beta$ controls the faint-end power-law behavior, $\gamma$ controls the curvature of the bright-end cutoff, and $\Phi_{\star}$ is the normalization.
Using $L/L_{\star}=10^{-0.4(M-M_{\star})}$, the corresponding local magnitude-space form entering Equation~(\ref{eq:shape_preserving_qlf_M}) is
\begin{eqnarray}
\label{eq:saunders_mag}
\begin{aligned}
\phi_{\rm Sau}(z=0,M)
=
\phi_{\star}
10^{-0.4(1-\beta)(M-M_{\star})}
\exp\left\{
-\frac{1}{2\gamma^2}
\left[
\log_{10}
\left(
1 + 10^{-0.4(M-M_{\star})}
\right)
\right]^2
\right\}.
\end{aligned}
\end{eqnarray}

The constant Jacobian factor between $d\log_{10}L$ and $dM$ is absorbed into the normalization $\phi_{\star}$.
Similarly, for the Saunders local LF, the characteristic magnitude evolves as
$M_{\star}(z)=M_{\star}-2.5\log_{10}e_2(z)$.
Compared with the DPL, the Saunders local LF replaces the asymptotic bright-end power law with a smooth log-normal-like cutoff.
In this work, we use this form as a controlled alternative local-shape family to test whether the preferred global LADE evolution depends on the assumed local LF shape.

\section{Full LADE model-grid rankings}
\label{app:full_model_grid}
\setcounter{table}{0}
\setcounter{figure}{0}
\setcounter{equation}{0}

Table~\ref{tab:full_model_grid} reports the complete LADE model-grid rankings for the two local-LF branches considered in this work.
For each local LF form, all combinations of the candidate density-evolution and luminosity-evolution functions listed in Table~\ref{tab:evolution_function_library} are fitted using the same likelihood, data selection, and survey selection functions as in the main analysis.
The DPL and Saunders local-LF branches are ranked independently by increasing AIC.
The column $q$ gives the total number of free parameters.
Models marked with $\dagger$ have both AIC and BIC values lower than those of the FDPL-M1 reference model.

In the DPL local-LF branch, the first twelve ranked LADE models satisfy this criterion, indicating that the preference over the FDPL reference fit is not confined to the single fiducial model.
In the Saunders local-LF branch, three LADE models also have both AIC and BIC values lower than those of the FDPL reference models.
This difference indicates that the DPL local LF provides a better baseline description of the UV QLF shape, while the existence of multiple preferred LADE models in both branches supports the robustness of the shape-preserving evolutionary framework.
In both local-LF branches, the best-ranked model adopts the same D6L8 density--luminosity evolution combination, further indicating that the preferred evolutionary structure is not driven solely by the adopted local LF form.

\begingroup
\scriptsize
\setlength{\LTleft}{3pt}
\setlength{\LTright}{3pt}
\setlength{\LTcapwidth}{\textwidth}
\setlength{\tabcolsep}{8pt}
\renewcommand{\arraystretch}{1.05}
\begin{longtable}{cclcc@{\hspace{1.0em}}!{\color{gray!25}\vrule width 0.5pt}@{\hspace{1.0em}}cclcc}
\caption{
Full AIC and BIC rankings for the LADE model-grid search.
}
\label{tab:full_model_grid}\\
\hline\hline
\multicolumn{5}{c}{DPL local LF} &
\multicolumn{5}{c}{Saunders local LF} \\
\cline{1-5}\cline{6-10}
Rank & $q$ & Model & AIC & BIC &
Rank & $q$ & Model & AIC & BIC \\
\hline
\endfirsthead
\caption{Continued.}\\
\hline\hline
\multicolumn{5}{c}{DPL local LF} &
\multicolumn{5}{c}{Saunders local LF} \\
\cline{1-5}\cline{6-10}
Rank & $q$ & Model & AIC & BIC &
Rank & $q$ & Model & AIC & BIC \\
\hline
\endhead
\hline
\multicolumn{10}{r}{\emph{Continued on next page}}\\
\endfoot
\hline\hline
\endlastfoot
1 & 10 & D6L8$^\dagger$ & 2173585.729 & 2173677.427 & 1 & 10 & D6L8$^\dagger$ & 2173770.119 & 2173861.817 \\
2 & 9 & D5L8$^\dagger$ & 2173749.379 & 2173831.908 & 2 & 10 & D9L8$^\dagger$ & 2173770.176 & 2173861.874 \\
3 & 10 & D9L9$^\dagger$ & 2173759.864 & 2173851.562 & 3 & 9 & D5L8$^\dagger$ & 2173914.006 & 2173996.535 \\
4 & 10 & D6L9$^\dagger$ & 2173759.956 & 2173851.654 & 4 & 10 & D9L9 & 2173946.535 & 2174038.233 \\
5 & 10 & D6L6$^\dagger$ & 2173761.191 & 2173852.889 & 5 & 10 & D9L6 & 2173946.536 & 2174038.234 \\
6 & 10 & D9L8$^\dagger$ & 2173764.484 & 2173856.182 & 6 & 10 & D6L9 & 2173946.632 & 2174038.330 \\
7 & 9 & D6L4$^\dagger$ & 2173862.051 & 2173944.580 & 7 & 10 & D6L6 & 2173983.780 & 2174075.478 \\
8 & 9 & D9L4$^\dagger$ & 2173862.083 & 2173944.612 & 8 & 9 & D6L4 & 2174066.073 & 2174148.601 \\
9 & 10 & D9L7$^\dagger$ & 2173873.780 & 2173965.478 & 9 & 9 & D9L4 & 2174066.130 & 2174148.658 \\
10 & 10 & D8L8$^\dagger$ & 2173879.948 & 2173971.646 & 10 & 10 & D9L7 & 2174066.644 & 2174158.342 \\
11 & 9 & D9L5$^\dagger$ & 2173899.895 & 2173982.423 & 11 & 9 & D6L5 & 2174091.931 & 2174174.459 \\
12 & 9 & D6L5$^\dagger$ & 2173899.923 & 2173982.451 & 12 & 9 & D9L5 & 2174091.956 & 2174174.484 \\
13 & 9 & D9L2 & 2173976.233 & 2174058.761 & 13 & 9 & D6L2 & 2174168.970 & 2174251.499 \\
14 & 9 & D6L3 & 2174082.127 & 2174164.656 & 14 & 9 & D9L2 & 2174168.972 & 2174251.500 \\
15 & 9 & D9L3 & 2174082.305 & 2174164.833 & 15 & 9 & D4L8 & 2174178.505 & 2174261.033 \\
16 & 8 & D9L1 & 2174091.615 & 2174164.973 & 16 & 9 & D6L3 & 2174286.521 & 2174369.049 \\
17 & 8 & D6L1 & 2174091.660 & 2174165.018 & 17 & 8 & D6L1 & 2174306.653 & 2174380.012 \\
18 & 10 & D6L7 & 2174160.351 & 2174252.050 & 18 & 8 & D9L1 & 2174306.695 & 2174380.053 \\
19 & 9 & D4L8 & 2174182.085 & 2174264.614 & 19 & 10 & D8L9 & 2174323.334 & 2174415.032 \\
20 & 9 & D8L5 & 2174237.292 & 2174319.821 & 20 & 10 & D8L7 & 2174435.506 & 2174527.204 \\
21 & 10 & D8L7 & 2174239.095 & 2174330.793 & 21 & 9 & D8L3 & 2174484.649 & 2174567.178 \\
22 & 9 & D8L2 & 2174252.068 & 2174334.596 & 22 & 8 & D8L1 & 2174497.459 & 2174570.817 \\
23 & 10 & D8L6 & 2174275.620 & 2174367.318 & 23 & 9 & D8L4 & 2174562.582 & 2174645.110 \\
24 & 9 & D8L3 & 2174280.577 & 2174363.105 & 24 & 9 & D9L3 & 2174564.089 & 2174646.618 \\
25 & 8 & D8L1 & 2174285.524 & 2174358.883 & 25 & 9 & D7L8 & 2174649.720 & 2174732.248 \\
26 & 9 & D7L8 & 2174398.977 & 2174481.505 & 26 & 9 & D3L8 & 2174730.495 & 2174813.023 \\
27 & 9 & D3L8 & 2174545.959 & 2174628.487 & 27 & 9 & D4L9 & 2174817.838 & 2174900.366 \\
28 & 9 & D5L6 & 2174600.058 & 2174682.586 & 28 & 9 & D4L6 & 2174818.434 & 2174900.962 \\
29 & 9 & D5L9 & 2174600.269 & 2174682.797 & 29 & 9 & D5L9 & 2174845.560 & 2174928.089 \\
30 & 9 & D4L6 & 2174607.382 & 2174689.911 & 30 & 9 & D5L6 & 2174846.009 & 2174928.537 \\
31 & 9 & D4L9 & 2174607.387 & 2174689.916 & 31 & 9 & D3L6 & 2174925.188 & 2175007.716 \\
32 & 9 & D3L9 & 2174705.731 & 2174788.259 & 32 & 9 & D3L9 & 2174925.260 & 2175007.788 \\
33 & 9 & D3L6 & 2174705.793 & 2174788.321 & 33 & 9 & D7L6 & 2174946.679 & 2175029.207 \\
34 & 8 & D1L6 & 2174729.664 & 2174803.023 & 34 & 9 & D7L9 & 2174946.713 & 2175029.242 \\
35 & 8 & D1L9 & 2174729.849 & 2174803.208 & 35 & 8 & D1L9 & 2174970.923 & 2175044.282 \\
36 & 10 & D8L9 & 2174733.882 & 2174825.580 & 36 & 8 & D1L6 & 2174970.930 & 2175044.289 \\
37 & 9 & D7L9 & 2174735.471 & 2174818.000 & 37 & 10 & D8L6 & 2174977.252 & 2175068.950 \\
38 & 9 & D7L6 & 2174735.478 & 2174818.007 & 38 & 8 & D2L6 & 2175257.118 & 2175330.477 \\
39 & 8 & D2L6 & 2174974.531 & 2175047.890 & 39 & 8 & D2L9 & 2175257.145 & 2175330.503 \\
40 & 8 & D2L9 & 2174974.639 & 2175047.998 & 40 & 8 & D4L4 & 2175741.189 & 2175814.547 \\
41 & 8 & D1L8 & 2175101.428 & 2175174.787 & 41 & 8 & D7L5 & 2175900.653 & 2175974.012 \\
42 & 8 & D4L4 & 2175580.051 & 2175653.410 & 42 & 9 & D7L7 & 2175900.899 & 2175983.427 \\
43 & 8 & D2L8 & 2175652.997 & 2175726.356 & 43 & 9 & D3L7 & 2175939.499 & 2176022.027 \\
44 & 8 & D7L5 & 2175731.078 & 2175804.436 & 44 & 8 & D3L2 & 2175949.628 & 2176022.987 \\
45 & 9 & D7L7 & 2175732.121 & 2175814.649 & 45 & 8 & D7L2 & 2175958.345 & 2176031.703 \\
46 & 8 & D4L3 & 2175754.356 & 2175827.714 & 46 & 9 & D4L7 & 2175968.155 & 2176050.683 \\
47 & 8 & D3L2 & 2175774.917 & 2175848.276 & 47 & 8 & D2L8 & 2175999.974 & 2176073.333 \\
48 & 8 & D3L5 & 2175783.380 & 2175856.739 & 48 & 8 & D4L5 & 2176002.777 & 2176076.135 \\
49 & 8 & D7L2 & 2175794.767 & 2175868.125 & 49 & 8 & D4L3 & 2176015.262 & 2176088.621 \\
50 & 9 & D4L7 & 2175822.242 & 2175904.771 & 50 & 8 & D3L5 & 2176027.357 & 2176100.715 \\
51 & 8 & D7L4 & 2175839.924 & 2175913.283 & 51 & 8 & D3L3 & 2176221.544 & 2176294.902 \\
52 & 8 & D4L5 & 2175860.765 & 2175934.124 & 52 & 8 & D7L4 & 2176234.602 & 2176307.961 \\
53 & 8 & D3L3 & 2176044.855 & 2176118.213 & 53 & 8 & D7L3 & 2176249.074 & 2176322.433 \\
54 & 8 & D7L3 & 2176081.521 & 2176154.879 & 54 & 8 & D4L2 & 2176269.588 & 2176342.947 \\
55 & 7 & D7L1 & 2176097.069 & 2176161.258 & 55 & 7 & D7L1 & 2176286.060 & 2176350.248 \\
56 & 8 & D5L4 & 2176135.051 & 2176208.410 & 56 & 7 & D3L1 & 2176415.768 & 2176479.957 \\
57 & 8 & D4L2 & 2176140.907 & 2176214.266 & 57 & 7 & D4L1 & 2176582.551 & 2176646.740 \\
58 & 7 & D3L1 & 2176211.888 & 2176276.077 & 58 & 8 & D3L4 & 2176624.793 & 2176698.151 \\
59 & 7 & D4L1 & 2176424.322 & 2176488.510 & 59 & 9 & D5L7 & 2176819.774 & 2176902.302 \\
60 & 9 & D5L7 & 2176582.587 & 2176665.115 & 60 & 8 & D5L5 & 2176862.169 & 2176935.528 \\
61 & 8 & D5L5 & 2176616.478 & 2176689.836 & 61 & 8 & D1L8 & 2177059.005 & 2177132.363 \\
62 & 8 & D1L7 & 2176994.140 & 2177067.499 & 62 & 8 & D5L2 & 2177300.783 & 2177374.141 \\
63 & 8 & D3L4 & 2176995.996 & 2177069.355 & 63 & 8 & D1L7 & 2177352.117 & 2177425.476 \\
64 & 8 & D5L2 & 2177040.746 & 2177114.105 & 64 & 7 & D1L5 & 2177438.726 & 2177502.914 \\
65 & 7 & D1L5 & 2177065.707 & 2177129.896 & 65 & 7 & D1L2 & 2177443.943 & 2177508.131 \\
66 & 7 & D1L2 & 2177098.232 & 2177162.421 & 66 & 9 & D8L5 & 2177446.108 & 2177528.637 \\
67 & 7 & D2L2 & 2177577.869 & 2177642.058 & 67 & 9 & D8L2 & 2177448.463 & 2177530.992 \\
68 & 8 & D2L7 & 2177579.665 & 2177653.023 & 68 & 7 & D2L2 & 2178111.147 & 2178175.336 \\
69 & 7 & D2L5 & 2177842.882 & 2177907.071 & 69 & 8 & D5L3 & 2178163.876 & 2178237.235 \\
70 & 9 & D6L2 & 2177940.243 & 2178022.771 & 70 & 8 & D2L7 & 2178173.582 & 2178246.941 \\
71 & 8 & D5L3 & 2177965.048 & 2178038.407 & 71 & 10 & D6L7 & 2178256.227 & 2178347.926 \\
72 & 7 & D1L4 & 2178375.255 & 2178439.444 & 72 & 7 & D2L5 & 2178512.444 & 2178576.633 \\
73 & 9 & D8L4 & 2178379.312 & 2178461.840 & 73 & 7 & D1L4 & 2178631.716 & 2178695.905 \\
74 & 7 & D5L1 & 2179599.681 & 2179663.870 & 74 & 7 & D5L1 & 2179741.225 & 2179805.414 \\
75 & 7 & D2L3 & 2179778.260 & 2179842.449 & 75 & 10 & D8L8 & 2179879.099 & 2179970.797 \\
76 & 7 & D1L3 & 2180055.828 & 2180120.017 & 76 & 7 & D2L3 & 2180169.315 & 2180233.503 \\
77 & 7 & D2L4 & 2180254.478 & 2180318.667 & 77 & 7 & D1L3 & 2180494.553 & 2180558.742 \\
78 & 6 & D1L1 & 2181026.306 & 2181081.325 & 78 & 7 & D2L4 & 2181119.789 & 2181183.977 \\
79 & 10 & D9L6 & 2181954.788 & 2182046.487 & 79 & 6 & D1L1 & 2181382.280 & 2181437.299 \\
80 & 9 & D3L7 & 2184605.376 & 2184687.905 & 80 & 8 & D5L4 & 2184900.140 & 2184973.498 \\
81 & 6 & D2L1 & 2235541.354 & 2235596.372 & 81 & 6 & D2L1 & 2237112.545 & 2237167.564 \\
\end{longtable}
\endgroup

\section{Posterior diagnostics and best-fitting parameters}
\label{app:posterior_diagnostics}
\setcounter{table}{0}
\setcounter{figure}{0}
\setcounter{equation}{0}

\begin{figure*}
        \centering
        \includegraphics[width=\textwidth]{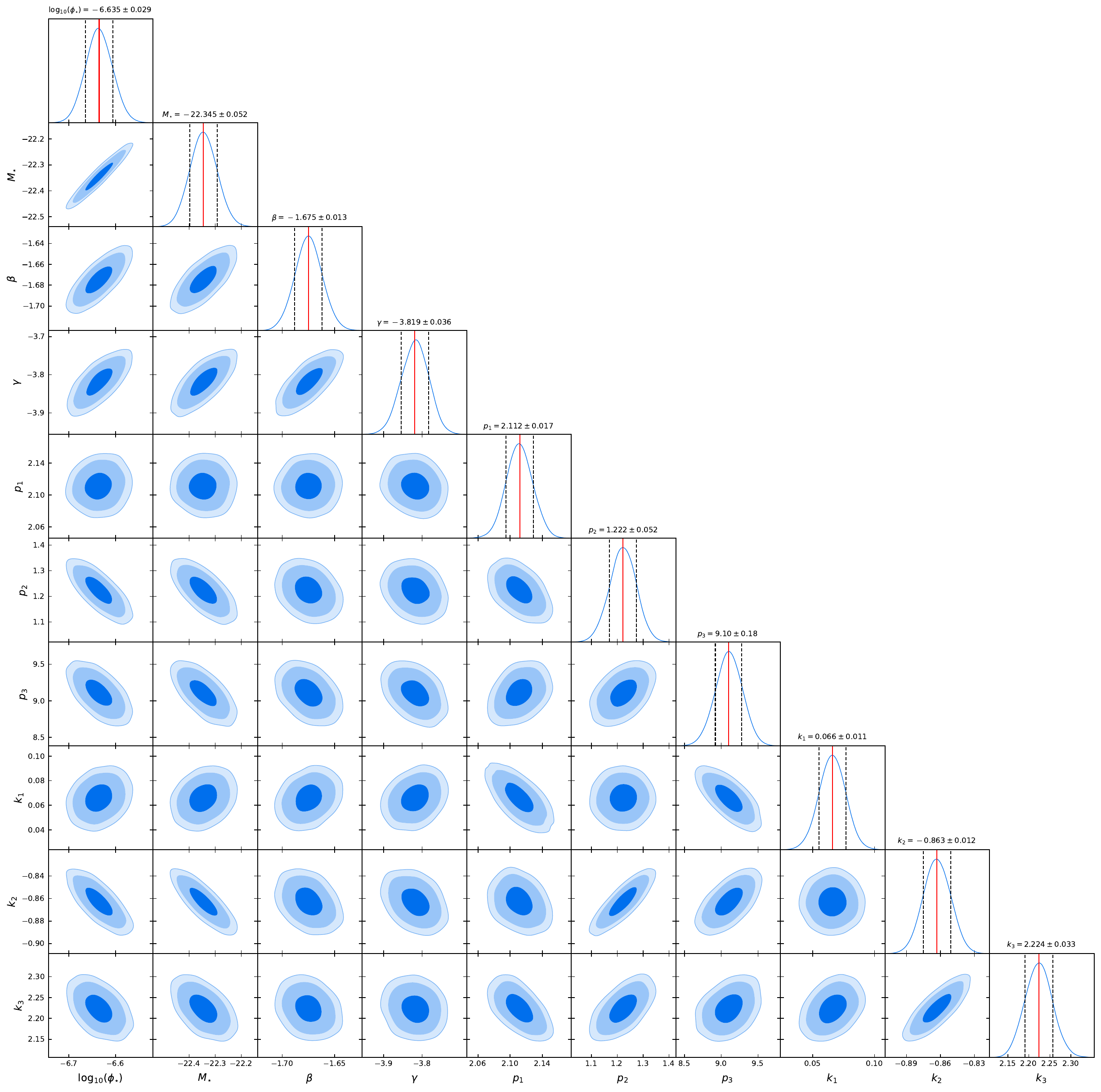}
        \caption{Corner plot illustrating the one- and two-dimensional projections of the posterior probability distributions for Model A, derived from the MCMC sampling. The diagonal panels display the marginalized posterior distributions for each parameter, with the 16th and 84th percentiles indicated by vertical dashed lines. The off-diagonal panels present the two-dimensional joint posterior distributions for each parameter pair, with 1$\sigma$, 2$\sigma$, and 3$\sigma$ confidence levels indicated by blue regions of varying depth. The red vertical solid lines indicate the best-fitting parameter values.}
        \label{fig:cornerplotA}
\end{figure*}

\begin{figure*}
        \centering
        \includegraphics[width=\textwidth]{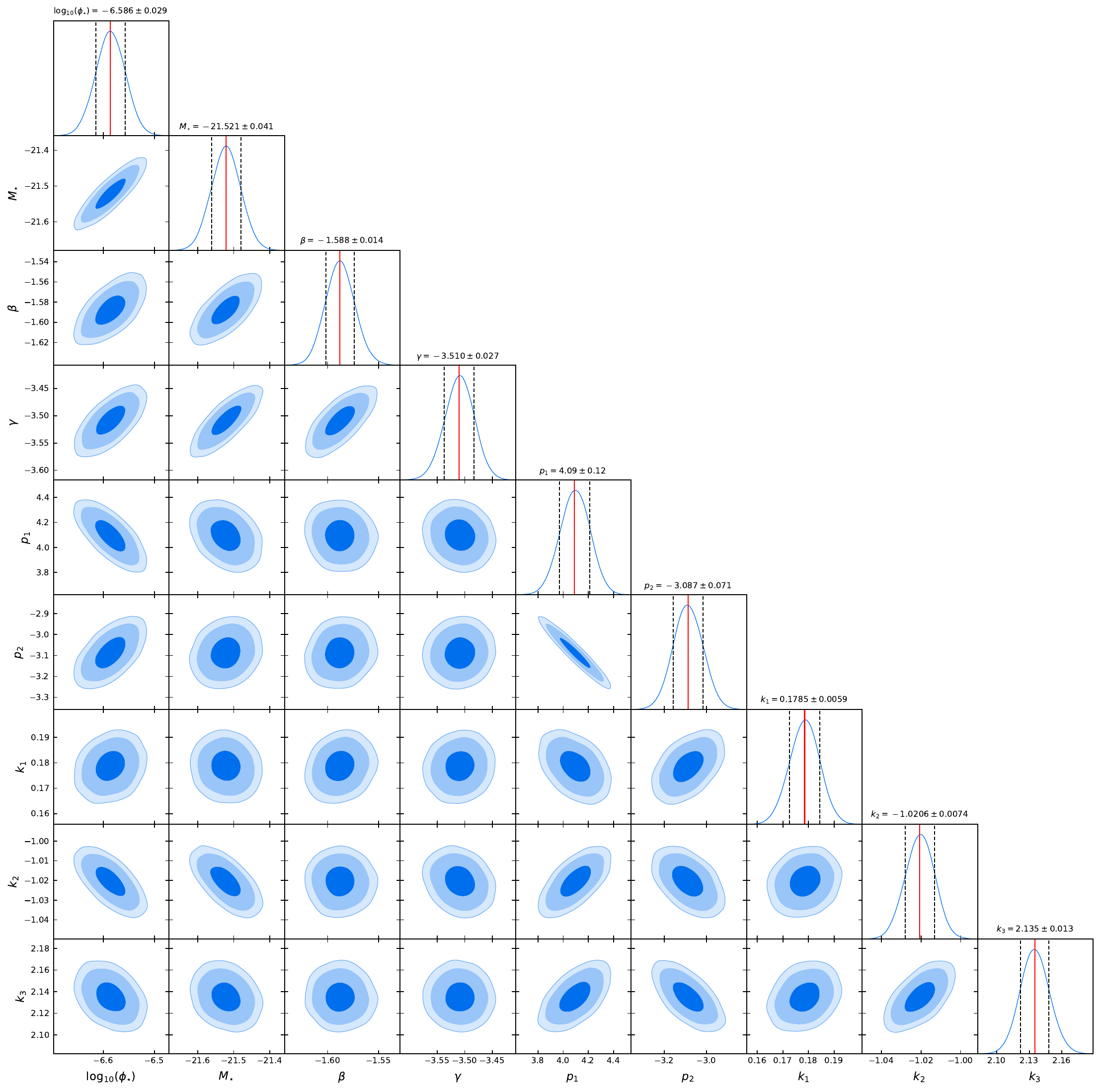}
        \caption{Similar to Figure \ref{fig:cornerplotA}, but for Model B.}
        \label{fig:cornerplotB}
\end{figure*}

\begin{figure*}
        \centering
        \includegraphics[width=\textwidth]{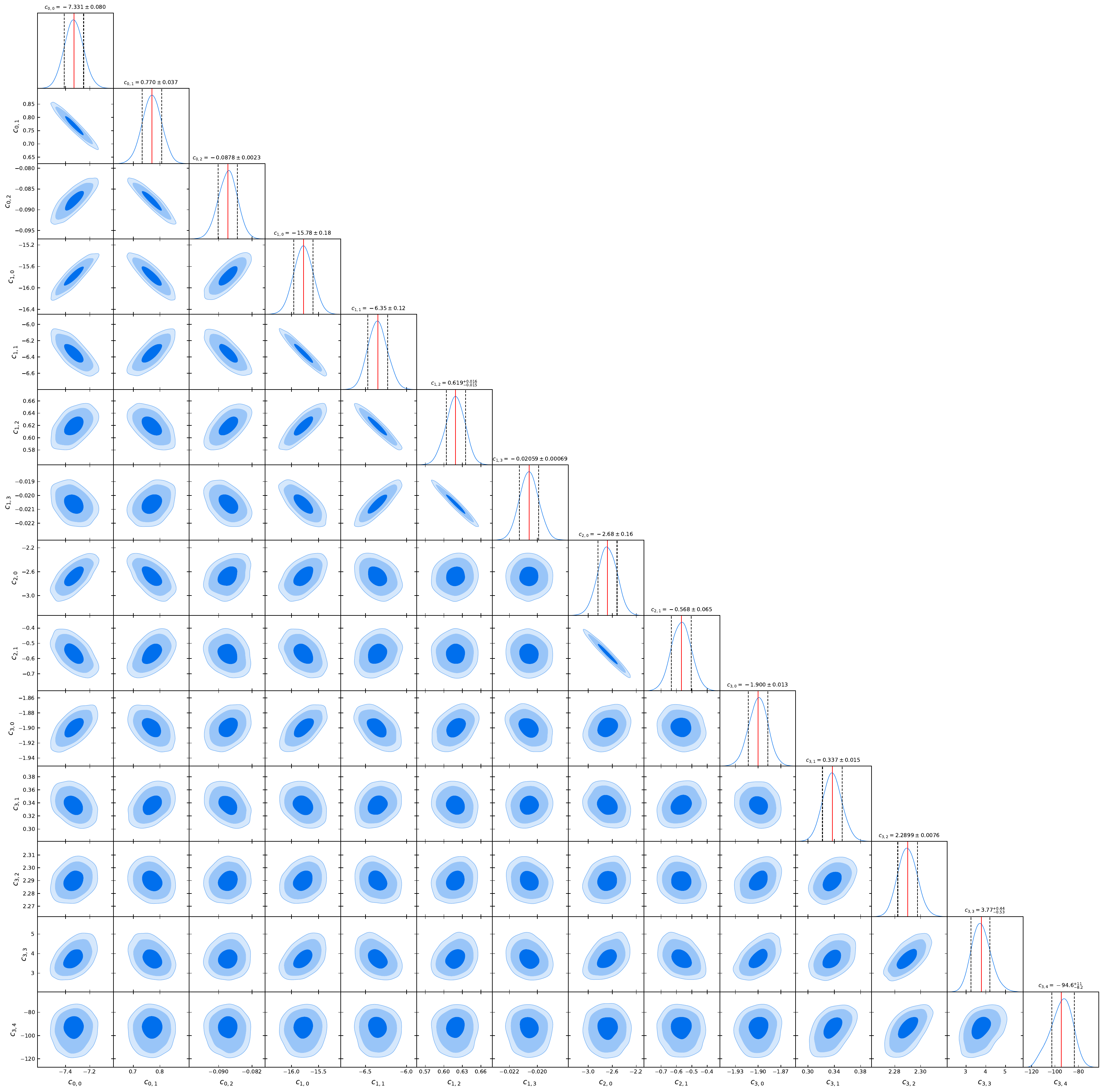}
        \caption{Similar to Figure \ref{fig:cornerplotA}, but for the FDPL-M1 reference model.}
        \label{fig:cornerplotChebyshev1}
\end{figure*}

\begin{figure*}
        \centering
        \includegraphics[width=\textwidth]{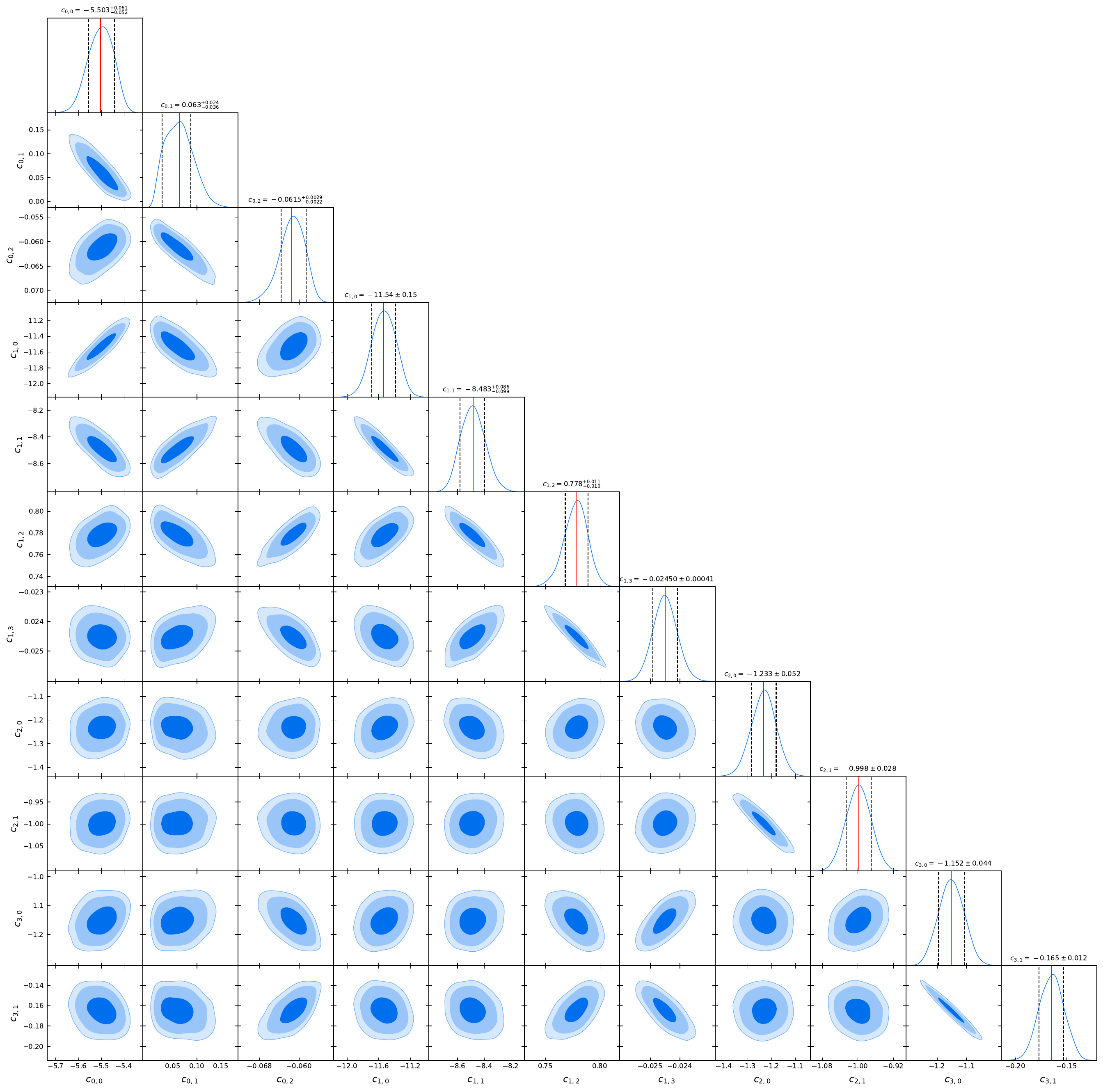}
        \caption{Similar to Figure \ref{fig:cornerplotA}, but for the FDPL-M3 reference model.}
        \label{fig:cornerplotChebyshev2}
\end{figure*}

This appendix provides posterior diagnostics for Models~A and B and for the FDPL reference models.
Figures~\ref{fig:cornerplotA} and \ref{fig:cornerplotB} show the marginalized posterior distributions for Models~A and B, whose best-fitting parameters are reported in Table~\ref{modelpara}.
Figures~\ref{fig:cornerplotChebyshev1}--\ref{fig:cornerplotChebyshev2}  and Table~\ref{modelpara_Chebyshev} give the corresponding posterior diagnostics and coefficients for the FDPL reference models.

\begin{table}[htbp!]
\centering
\caption{
Best-fitting coefficients for the FDPL reference models. The uncertainties are $1\sigma$ marginalized credible intervals.
}
\renewcommand{\arraystretch}{1.4}
\setlength{\tabcolsep}{3.5pt}
\begin{tabular}{lcccccccc}
\hline\hline
Model & $c_{0,0}$ & $c_{0,1}$ & $c_{0,2}$ & $c_{1,0}$ & $c_{1,1}$ & $c_{1,2}$ & $c_{1,3}$ \\
\hline
\raisebox{-0.4\height}{\shortstack[l]{FDPL-M1}}
& $-7.331_{-0.080}^{+0.080}$ & $0.770_{-0.037}^{+0.037}$
& $-0.088_{-0.002}^{+0.002}$ & $-15.780_{-0.180}^{+0.180}$
& $-6.350_{-0.120}^{+0.120}$ & $0.619_{-0.015}^{+0.016}$ & $0.021_{-0.001}^{+0.001}$ \\[7pt]
\raisebox{-0.4\height}{\shortstack[l]{FDPL-M3}}
& $-5.503_{-0.052}^{+0.061}$ & $0.063_{-0.036}^{+0.024}$
& $-0.062_{-0.002}^{+0.003}$ & $-11.54_{-0.150}^{+0.150}$
& $-8.483_{-0.099}^{+0.086}$ & $0.778_{-0.010}^{+0.011}$ & $-0.025_{-0.0004}^{+0.0004}$ \\[7pt]
\hline\hline

Model & $c_{2,0}$ & $c_{2,1}$ & $c_{3,0}$ & $c_{3,1}$ & $c_{3,2}$ & $c_{3,3}$ & $c_{3,4}$ \\
\hline

\raisebox{-0.4\height}{\shortstack[l]{FDPL-M1}}
& $-2.680_{-0.160}^{+0.160}$ & $-0.568_{-0.065}^{+0.065}$
& $-1.900_{-0.013}^{+0.013}$ & $0.337_{-0.015}^{+0.015}$
& $2.290_{-0.008}^{+0.008}$ & $3.770_{-0.530}^{+0.440}$ & $-94.600_{-8.200}^{+11.000}$ \\[5pt]
\raisebox{-0.4\height}{\shortstack[l]{FDPL-M3}}
& $-1.233_{-0.052}^{+0.052}$ & $-0.998_{-0.028}^{+0.028}$
& $-1.152_{-0.044}^{+0.044}$ & $-0.165_{-0.012}^{+0.012}$
& $\cdot\cdot\cdot$ & $\cdot\cdot\cdot$ & $\cdot\cdot\cdot$ \\[5pt]
\hline
\end{tabular}

\label{modelpara_Chebyshev}
\end{table}

\end{document}